\documentclass[twocolumn]{aastex63}

\shorttitle{Transients in TDE-Preferred Hosts}
\shortauthors{Arcavi, Nyiha \& French}

\graphicspath{{./}}
\usepackage{longtable}

\defcitealias{French2018}{FZ18}

\begin{document}

\title{Types of Transients in the Centers of Post-starburst and Quiescent Balmer-strong Galaxies}

\author[0000-0001-7090-4898]{Iair Arcavi}
\affiliation{The School of Physics and Astronomy, Tel Aviv University, Tel Aviv 69978, Israel}
\affiliation{CIFAR Azrieli Global Scholars program, CIFAR, Toronto, Canada}

\author[0000-0001-7187-6251]{Irura Nyiha}
\affiliation{Massachusetts Institute of Technology, Cambridge, MA 02139, USA}

\author[0000-0002-4235-7337]{K. Decker French}
\affiliation{Department of Astronomy, University of Illinois, 1002 W. Green St., Urbana, IL 61801, USA}
\affiliation{Center for Astrophysical Surveys, National Center for Supercomputing Applications, Urbana, IL, 61801, USA}

\correspondingauthor{Iair Arcavi}
\email{arcavi@tauex.tau.ac.il}

\begin{abstract}
Tidal Disruption Events (TDEs) have been found to show a preference for post-starburst (PS) and quiescent Balmer-strong (QBS) galaxies. This preference can be used to help find TDEs in transient surveys. But what other transients might ``contaminate'' such a search, and by how much? We examine all reported transients coincident with the centers of galaxies in the \cite{French2018} catalog of spectroscopically confirmed PS and QBS galaxies and photometrically identified PS and QBS galaxy candidates. We find that TDEs and Type Ia supernovae (SNe) are the only types of transients classified in the centers of these galaxies (aside from one active galactic nucleus flare), with Type Ia SNe being $8.3\pm0.2$ times more prevalent than TDEs ($1\sigma$ confidence bounds). This factor is $\sim$2.7 times lower than in a control sample of quiescent galaxies. Narrowing the sample to spectroscopically confirmed QBS galaxies does not change these statistics much. In spectroscopically confirmed PS galaxies, however, TDEs are the ones that outnumber Type Ia SNe $2\pm0.6$ to $1$. Unfortunately, there are few such galaxies in the catalog. By classifying transients from the entire catalog, three times more TDEs are expected to be found, but with a $\sim$16-times larger Type Ia SN contamination. We use the public ZTF photometric archive to search for possibly missed TDEs in the \cite{French2018} galaxies. We find three unclassified clear transients -- none of which are likely missed TDEs based on their light-curve colors.

\end{abstract}

\keywords{E+A galaxies (424), Supernovae (1668), Tidal disruption (1696)}

\section{Introduction}

Tidal disruption events (TDEs), caused when a star is torn apart by tidal forces around a supermassive black hole \citep{Rees1988}, can generate observable flares. Such events are rare \citep[e.g.][]{Wang2004,Stone2016}, but they are unique tools for learning about the population of otherwise quiescent supermassive black holes, accretion physics, strong gravity, and more. Therefore, finding them in transient surveys is desirable. However, such surveys are already producing orders of magnitude more transient candidates than can be vetted and classified spectroscopically. Any preference of TDEs for specific host galaxy types could be used not only to learn about the dynamical processes driving TDE rates, but also to help narrow the search for such events. 

\begin{table*}[t]
    \centering
    \caption{\label{tab:sources}Sources of the Galaxies Consolidated from \citetalias{French2018}.}
    \begin{tabular}{llll}
        \hline
         FZ18 & Source Name & No. of & No. of Unique \\
         Table No. & & Galaxies & Galaxies$^a$ \\
         \hline
         \hline
         1 & Spectroscopically Identified QBS Galaxies from SDSS & 19,514 & 19,514\\
         2 & Spectroscopically Identified PS Galaxies from SDSS & 1683 & 50$^b$ \\
         \hline
         5 & Pan-STARRS + WISE Photometrically Identified QBS Galaxies & 57,299 & 57,254 \\
         6 & DES + WISE Photometrically Identified QBS Galaxies & 9337 & 9296 \\
         7 & SDSS + WISE Photometrically Identified QBS Galaxies & 848 & 832\\
         \hline
         8 & Pan-STARRS + WISE Photometrically Identified PS Galaxies & 9690 & 750 \\
         9 & DES + WISE Photometrically Identified PS Galaxies & 753 & 44\\
         10 & SDSS + WISE Photometrically Identified PS Galaxies & 117 & 8 \\
         \hline
         Total & & & 87,748 \\
         \hline
    \end{tabular}
    \tablecomments{Of these, we removed 53 objects from Tables 5--10 given SDSS DR16 spectra indicating they are QSO's or stars.\\
    $^a$Number of galaxies not included in the sources from the previous rows.\\
    $^b$Omitted from the source above it by mistake, here we include these galaxies also in Source 1.}
\end{table*}

\cite{Arcavi2014} discovered that optical TDEs occur preferentially in post-starburst (PS; also known as ``E+A'') galaxies. \cite{French2016} later quantified this preference, expanding the definition of the preferred hosts to include also quiescent Balmer-strong (QBS) galaxies. The reason that TDEs prefer PS and QBS galaxies is not yet fully understood (see \citealt{French2020} for a recent review); however it can still be leveraged to help identify promising TDE candidates in transient surveys.

\citet[][hereafter FZ18]{French2018} define QBSs as having a Lick H$\delta_A$ index $>1.3\,{\textrm \AA}$ in absorption and an H$\alpha$ equivalent width $<5\,{\textrm \AA}$ in emission. The PS galaxies are a subset of these, defined with H$\delta_A$ $>4\,{\textrm \AA}$ in absorption and H$\alpha$ equivalent width $<3\,{\textrm \AA}$ in emission. 

Unfortunately, spectra are not available for most galaxies. Thus, \citetalias{French2018} use spectroscopically confirmed QBS and PS galaxies from the Sloan Digital Sky Survey \citep[SDSS;][]{York2000} Data Release (DR) 12 main galaxy survey \citep{Strauss2002,Alam2015} to train a machine-learning algorithm to identify QBS and PS galaxies from photometry alone. They then run this algorithm on a combination of Panoramic Survey Telescope and Rapid Response System \citep[Pan-STARRS;][]{Chambers2016} and Wide-field Infrared Survey Explorer \citep[WISE;][]{Wright2010} data, Dark Energy Survey \citep[DES;][]{Abbott2018} and WISE data, and SDSS and WISE data to identify several tens of thousands of new QBS and PS galaxy candidates.

Here, we search the the Transient Name Server (TNS)\footnote{\url{http://www.wis-tns.org}} database and the Zwicky Transient Facility \citep[ZTF;][]{Bellm2014,Graham2019} public photometric data for transients coincident with the centers of galaxies in the \citetalias{French2018} catalog. Our goals are to (1) measure the relative observed fractions of different types of transients occurring in the centers of such galaxies from the TNS data, and (2) check if any unclassified transients in these galaxies could have been missed TDEs, using the ZTF photometric data.

We adopt the nine-year Wilkinson Microwave Anisotropy Probe (WMAP) cosmology \citep{Hinshaw2013} throughout. 

\section{Consolidating the FZ18 Galaxy Catalog}

The \citetalias{French2018} catalog is divided into eight subcatalogs, depending on how each galaxy was selected (hereafter we refer to these subcatalogs as ``sources"). We number each source according to its table number in \citetalias{French2018} and list them here in Table \ref{tab:sources}.

\begin{figure}
\includegraphics[width=\columnwidth]{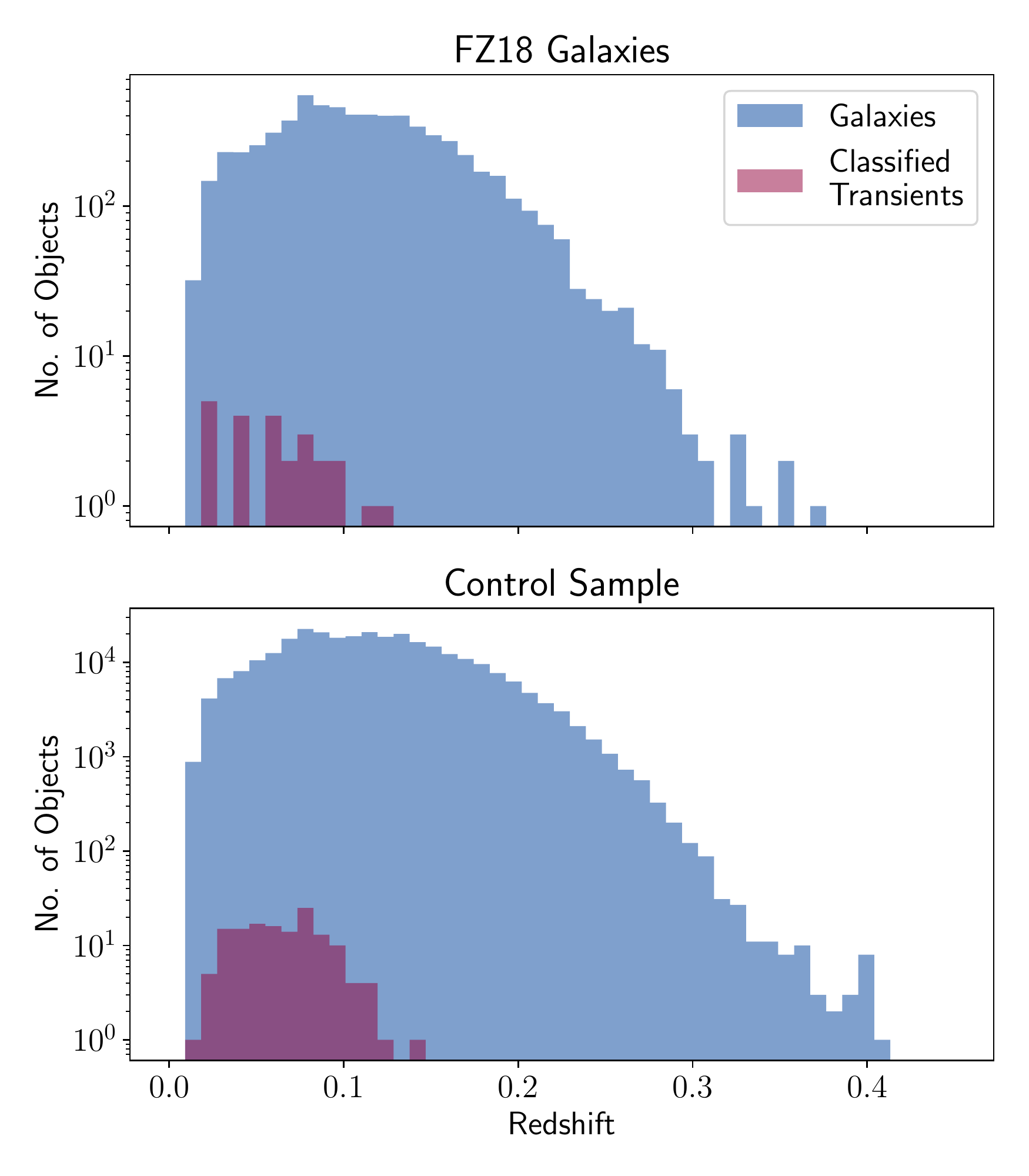}
\caption{\label{fig:redshifts}Redshift distributions of the galaxies and classified transients coincident with their centers for the \citetalias{French2018} catalog (top; only galaxies from Sources 1 and 2 are shown) and the control catalog (bottom).} 
\end{figure}

We consolidate the galaxies identified by \citetalias{French2018} from all sources into one master catalog. Since there is some overlap between galaxies in different sources, we note in the last column of Table \ref{tab:sources} the number of new galaxies in each source that were not already included in the sources from the previous rows. In total, we obtain 87,748 unique galaxies\footnote{We find 50 galaxies in Source 2 (spectroscopically identified PS galaxies) that are not in Source 1 (spectroscopically identified QBS galaxies), even though Source 2 should be a subset of Source 1. Indeed, these galaxies were omitted from Source 1 in \citetalias{French2018} by mistake. Here, we include them also as members of Source 1 for the rest of the analysis.}. The redshift distribution of the spectroscopically identified galaxies (Sources 1 and 2) is plotted in the top panel of Figure \ref{fig:redshifts}.

Since the \citetalias{French2018} catalog was compiled from SDSS DR12 data, we check which galaxies in the photometrically selected catalog (i.e. sources 5-10) of \citetalias{French2018} have since been observed spectroscopically by SDSS in DR16. We find that 3309 galaxies in the photometrically selected catalog have SDSS DR16 spectra. Of these, 41 have a ``Quasi Stellar Object'' (QSO) classification, and 12 have a ``Star'' classification (the rest all have a ``Galaxy'' classification). We remove from our sample the 53 galaxies with a spectrum having either a ``QSO'' or ``Star'' classification. Our final galaxy catalog thus consists of 87,695 galaxies.

\section{The Control Galaxy Catalog}

For a control sample we compile a catalog of quiescent galaxies (which are not necessarily Balmer-strong). We select all SDSS DR16 spectroscopically observed galaxies with an H$\alpha$ equivalent width $<3\,{\textrm \AA}$ in emission, as used for the \citetalias{French2018} PS cut (i.e. Source 2). We also require (as done in \citetalias{French2018}) that the redshift of each galaxy be $>0.01$ to avoid aperture bias, that the median signal to noise ratio of the spectrum be $>10$, and that the \texttt{h\_alpha\_eqw\_err} parameter be $>-1$ (i.e. no error flags were reported in the equivalent-width measurement).
These are the same cuts used for the \citetalias{French2018} Source 2 galaxies, just without the H$\delta_A$ absorption requirement. 

We find 297,284 such galaxies, which we designate as our control sample (of these, 13,213 are also in the \citetalias{French2018} catalog). Their redshift distribution is very similar to that of the spectroscopically identified \citetalias{French2018} galaxies, and is shown in the bottom panel of Figure \ref{fig:redshifts}.   
  
\section{Searching for Transients}

\begin{table*}[t]
\caption{\label{tab:tnsfracs}Number of Transients in the TNS (and Their Reported TNS Classifications) within 1\arcsec\ of a Galaxy in the Control Sample, in the \citetalias{French2018} Catalogs, and in Each of Its Subcatalogs (or ``Sources'').}
\begin{tabular}{lllllll}
\hline 
\hline
Source & Total & Not & SN Ia & TDE & AGN & Galaxy \\
& Transients & Classified & & & \\
\hline
\multicolumn{7}{c}{Control Sample} \\
\hline
Control Catalog & 726 & 577 & 136 & 6 & 1 & 6 \\
Percentage of All Transients &  & 79\% & 19\% & 1\% & 0\% & 1\% \\
Percentage of Classified Transients &  &  & 91\% & 4\% & 1\% & 4\% \\
\hline
\multicolumn{7}{c}{FZ18 Catalog} \\
\hline
All FZ18 Galaxies & 101 & 71 & 25 & 3 & 1 & 1 \\
Percentage of All Transients &  & 70\% & 25\% & 3\% & 1\% & 1\% \\
Percentage of Classified Transients &  &  & 83\% & 10\% & 3\% & 3\% \\
\hline
\hspace{0.2cm}1: SDSS Spec Identified QBSs & 74 & 50 & 20 & 3 & 0 & 1 \\
\hspace{0.2cm}Percentage of All Transients &  & 68\% & 27\% & 4\% & 0 & 1\% \\
\hspace{0.2cm}Percentage of Classified Transients &  &  & 83\% & 12\% & 0 & 4\% \\
\hline
\hspace{0.2cm}2: SDSS Spec Identified PSs & 10 & 7 & 1 & 2 & 0 & 0 \\
\hspace{0.2cm}Percentage of All Transients &  & 70\% & 10\% & 20\% & 0 & 0 \\
\hspace{0.2cm}Percentage of Classified Transients &  &  & 33\% & 67\% & 0 & 0 \\
\hline
\hspace{0.2cm}5: Pan-STARRS+WISE Phot Identified QBSs & 22 & 17 & 4 & 0 & 1 & 0 \\
\hspace{0.2cm}Percentage of All Transients &  & 77\% & 18\% & 0 & 5\% & 0 \\
\hspace{0.2cm}Percentage of Classified Transients &  &  & 80\% & 0 & 20\% & 0 \\
\hline
\hspace{0.2cm}6: DES+WISE Phot Identified QBSs & 2 & 2 & 0 & 0 & 0 & 0 \\
\hspace{0.2cm}Percentage of All Transients &  & 100\% & 0 & 0 & 0 & 0 \\
\hspace{0.2cm}Percentage of Classified Transients &  &  & 0 & 0 & 0 & 0 \\
\hline
\hspace{0.2cm}7: SDSS+WISE Phot Identified QBSs & 3 & 2 & 1 & 0 & 0 & 0 \\
\hspace{0.2cm}Percentage of All Transients &  & 67\% & 33\% & 0 & 0 & 0 \\
\hspace{0.2cm}Percentage of Classified Transients &  &  & 100\% & 0 & 0 & 0 \\

%\hline
%8: Pan-STARRS+WISE Phot-Identified PS & 0 & 0 & 0 & 0 & 0 & 0 \\
%Percentage of All Transients &   & 0 & 0 & 0 & 0 & 0 \\
%Percentage of Classified Transients &   &   & 0 & 0 & 0 & 0 \\
%\hline
%9: DES+WISE Phot-Identified PS & 0 & 0 & 0 & 0 & 0 & 0 \\
%Percentage of All Transients &   & 0 & 0 & 0 & 0 & 0 \\
%Percentage of Classified Transients &   &   & 0 & 0 & 0 & 0 \\
%\hline
%10: SDSS+WISE Phot-Identified PS & 0 & 0 & 0 & 0 & 0 & 0 \\
%Percentage of All Transients &   & 0 & 0 & 0 & 0 & 0 \\
%Percentage of Classified Transients &   &   & 0 & 0 & 0 & 0 \\

\hline
\hline
\end{tabular}
\tablecomments{No events were found in sources 8,9 and 10, so they are omitted from the table.}
\end{table*}

We next perform an archival search for transients coincident to within 1\arcsec\ with the centers of galaxies in both the \citetalias{French2018} and the control catalogs. This angular-separation cut is used to account for possible inaccuracies in transient or galaxy localizations. It corresponds to $\sim$2 kiloparsecs at the mean galaxy redshift of the \citetalias{French2018} spectroscopic sample, and $\sim$1 kiloparsec at the mean transient redshift of matched events.

\subsection{TNS Search}

\begin{figure*}[t]
\centering
\includegraphics[width=0.7\textwidth]{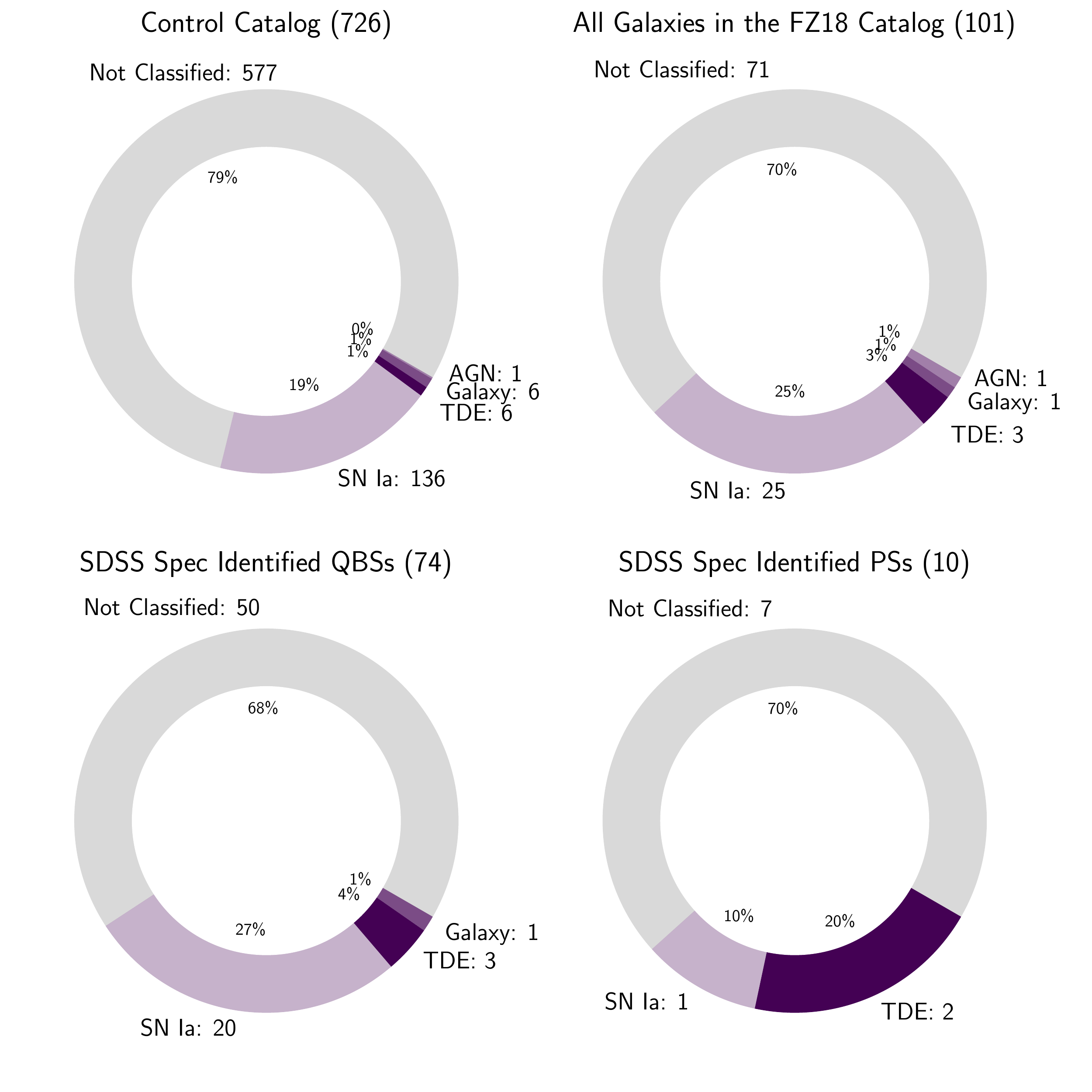}\includegraphics[width=0.3\textwidth]{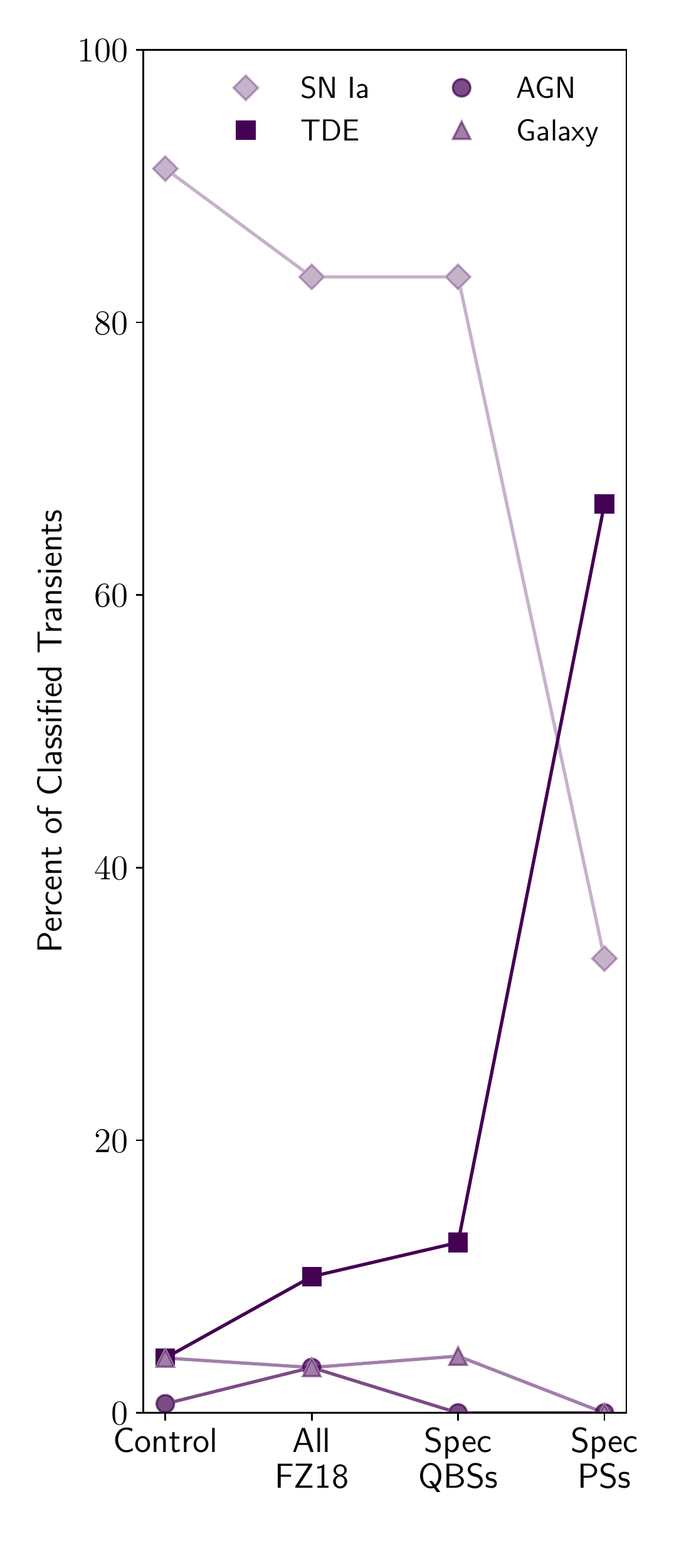}
\caption{\label{fig:classpie}TNS classifications of transients coincident with the centers of galaxies in the control sample, the entire \citetalias{French2018} galaxy catalog, and in the spectroscopically identified QBS and PS galaxies. The share of TDEs among classified transients is larger in the \citetalias{French2018} catalog compared to the control catalog. Specifically, in spectroscopically identified PS galaxies, most classified transients turn out to be TDEs (though the number of transients there is small).} 
\end{figure*}

TNS is the official International Astronomical Union (IAU) service for reporting transient events. It incorporates all events reported to the IAU through circulars before the TNS existed in its current form. Public spectroscopic classifications of transients are also reported to the TNS.

We search the complete TNS database up to 2021 August 8 for transients with positions within 1\arcsec\ of objects in the \citetalias{French2018} catalog. We find 101 such objects\footnote{Of these, 37 were identified by ZTF, 18 by ATLAS, 10 by Pan-STARRS, 3 by iPTF, 3 by Gaia, and 2 by ASAS-SN.} (Table \ref{tab:tns} in the Appendix), of which 30\% are spectroscopically classified (their redshift distribution is shown in the top panel of Figure \ref{fig:redshifts}). Of those, 83\% are Type Ia SNe\footnote{Here, we do not distinguish between the different subtypes of SNe Ia.}, and 10\% are TDEs. One event was an active galactic nucleus (AGN) flare and one is classified as ``Galaxy'' (i.e. only galaxy light was visible in the classification spectrum). This could mean that the event was not real, or that it faded before the spectrum was obtained. In the latter case, it could be a missed, rapidly evolving transient, hence we keep it in the sample. 

All TDEs are found within 0\farcs5 of their host, while Type Ia SNe show a uniform host-offset distribution out to our cut of 1\arcsec. A cut of 0\farcs5 would decrease the Type Ia SN fraction to 70\% and increase the TDE fraction to 15\%. However, we are dealing with small absolute numbers. A larger sample is required to more accurately analyze class fraction trends with measured host separations. Here, we keep the cut at 1\arcsec\ to avoid biases related to position measurement accuracy.

In the control catalog of quiescent galaxies we find 726 transients coincident to within 1\arcsec\ of a galaxy. Here, only 20\% of the transients are spectroscopically classified (their redshift distribution is shown in the bottom panel of Figure \ref{fig:redshifts}). Of those, 91\% are Type Ia SNe, and only 4\% are TDEs (half of which are in galaxies included also in the \citetalias{French2018} catalog). The rest are classified as AGN or ``Galaxy''\footnote{One event, SN 2018aii, has an ambiguous classification as either a Type Ia or Type Ic SN. Given the quiescent host galaxy, it is more likely to be a Type Ia SN, and we thus include it in that count. In any case, either option has a negligible effect on our statistical results.}.

If we remove from the control sample the 13,213 quiescent Balmer-strong galaxies that are also in the \citetalias{French2018} sample, we find 669 transients, of which 19\% are spectroscopically classified. Of those, 93\% are Type Ia SNe. Because half of the TDEs in the quiescent sample are also in the quiescent Balmer-strong sample, removing them lowers the fraction of spectroscopically confirmed TDEs even further to 2\%. The rest of the classified transients are AGN or ``Galaxy'', as in the full control catalog.

Table \ref{tab:tnsfracs} lists the number of events and their classifications for the full control catalog, the entire \citetalias{French2018} catalog, and per the \citetalias{French2018} sources (no transients were reported in the centers of galaxies from Sources 8--10). Figure \ref{fig:classpie} presents the distribution of classes of transients coincident with the center of a galaxy in the control sample, the entire \citetalias{French2018} catalog, and for transients only in the spectroscopically identified PS and QBS galaxies. 

\subsection{ZTF Search}

We next search the ZTF public alert stream for transient candidates with positions within 1\arcsec\ of objects in the \citetalias{French2018} galaxy catalog to check for any possible missed TDEs that were not reported to the TNS or were not classified there. We do this by using the ``E+A Galaxies'' watchlist\footnote{\url{https://lasair.roe.ac.uk/watchlist/321/}} on the Lasair Broker \citep{Smith2019}. 

We find 395 ZTF events as of 2021 August 8 coincident with a galaxy in the \citetalias{French2018} catalog (Table \ref{tab:ztf})\footnote{Here we removed 7 events which are in the Lasair watchlist, but are in galaxies identified by SDSS DR16 as ``QSO'' or ``Star''.}. 
Of those, 69 were reported to the TNS (and are therefore also included in Table \ref{tab:tns}), and 25 have classifications on the TNS. 

\begin{figure}
\includegraphics[width=\columnwidth]{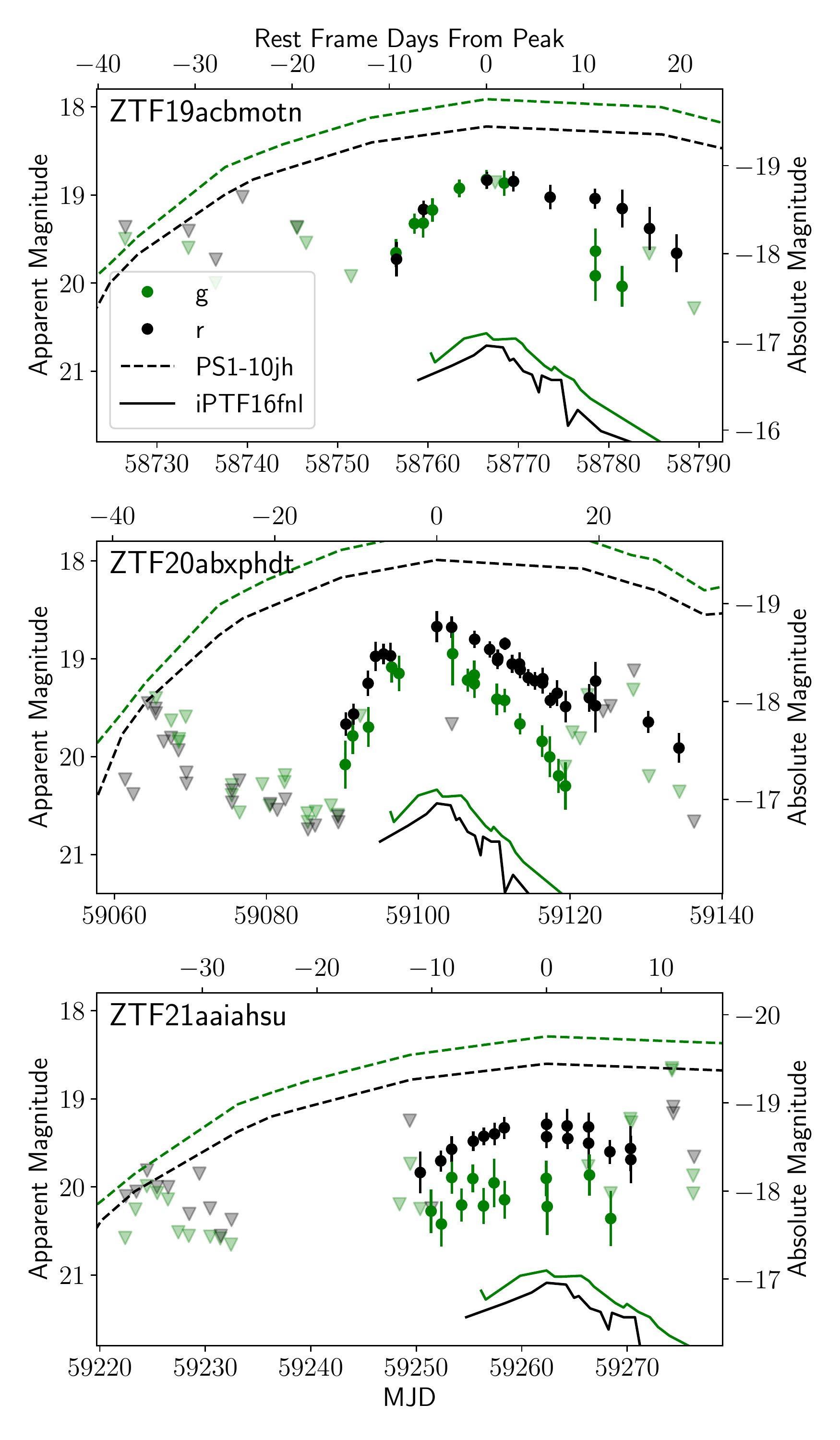}
\caption{\label{fig:gold}ZTF light curves of our ``Gold'' set of unclassified transients coincident with the center of a galaxy in the \citetalias{French2018} catalog (triangles denote 5$\sigma$ nondetection upper limits). Each light curve is compared to those of two optical TDEs: the prototypical PS1-10jh \citep{Gezari2012} and the rapidly evolving iPTF16fnl \citep{Blagorodnova2017,Brown2018}. The comparison light curves are aligned to the top (rest-frame days from peak) and right (absolute magnitude) axes of each plot. While all events have plausible time scales and luminosities to be TDEs, all are redder than the reference TDEs.} 
\end{figure}

We wish to check for missed TDEs among the unclassified events using their publicly available light curves. To do this, we retrieve the ZTF photometry of unclassified events with at least 20 detections, using the ALeRCE broker \citep{Forster2021} client\footnote{\url{https://alerce.readthedocs.io/en/latest/}}. We divide the light curves qualitatively into three groups: ``Gold'' -- those that are clearly transient showing a coherent rise and fall (three objects; Fig. \ref{fig:gold}), ``Silver'' -- those that are clearly variable (one object; Fig. \ref{fig:silver}), and ``Bronze'' -- those showing a rise and then remaining constant, or showing upper limits intertwined with detections, indicating they might be subtraction artifacts or flaring Galactic sources (27 objects; Fig. \ref{fig:bronze}). 

\section{Analysis and Discussion}

\subsection{TNS Search: Distribution of Transients Classes}

We use the Clopper–-Pearson method \citep{clopper1934} to estimate the confidence bounds of the observed ratios. This method uses binomial statistics to estimate lower and upper confidence bounds for ratios of rates of different 
event types, when the numbers of observed 
events are small \citep[as discussed by][]{Gehrels1986}. 

Using the $1\sigma$ confidence bounds calculated with this method, we find that in the control sample of quiescent galaxies, $91.3\%\pm2.3\%$ of transients are classified as Type Ia SNe, and only $4.0\%\pm1.6\%$ are classified as TDEs. Thus, one should expect to find $22.7\%\pm0.1$ times more Type Ia SNe than TDEs in such galaxies. In the \citetalias{French2018} catalog, by contrast, the Type Ia SN prevalence decreases to $83.3\%\pm6.8\%$ and the TDE prevalence increases to $10.0\%\pm5.5\%$, decreasing the Type Ia SN to TDE ratio to $8.3\pm0.2$. 

\begin{figure}
\includegraphics[width=\columnwidth]{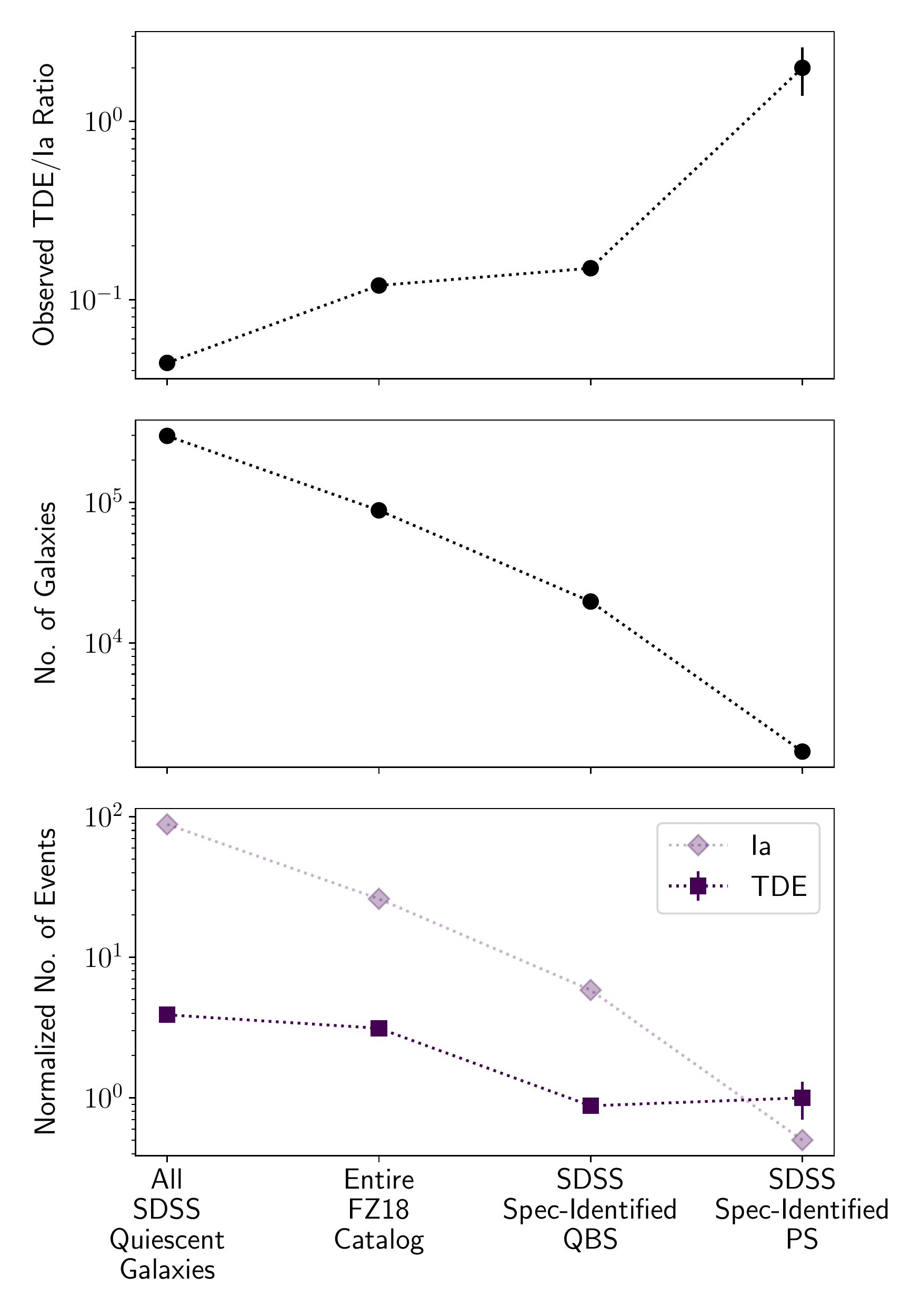}
\caption{\label{fig:no_of_events}Top: observed TDE to Type Ia SN ratio of transients in the centers of galaxies drawn from different galaxy catalogs analyzed here (1$\sigma$ Clopper--Pearson confidence bounds are shown but are sometimes smaller than the marker size). Middle: number of galaxies in each catalog. Bottom: total number of TDEs and Type Ia SNe expected in each galaxy catalog, normalized to the number of TDEs in spectroscopically identified PS galaxies. While the ratio of TDEs to Type Ia SNe is largest there, the small number of such galaxies in the catalog means that in absolute numbers, more TDEs can be discovered by using the entire \citetalias{French2018} catalog, but at the price of having $\sim$16 times more Type Ia SNe per TDE.} 
\end{figure}

This ratio is roughly the same when considering only spectroscopically confirmed QBS galaxies (rather than the entire \citetalias{French2018} catalog), but drastically improves for TDEs in spectroscopically confirmed PS galaxies. There, TDEs are $66.7\%\pm27.1\%$ of classified transients (with the rest being Type Ia SNe). The observed TDE to Type Ia SN ratio in these galaxies is thus $2.0\pm0.6$ to $1$. The ratio of TDEs to Type Ia SNe for these various samples is summarized in the top panel of Figure \ref{fig:no_of_events}.

These are not intrinsic transient rates in each galaxy type, but rather observed fractions. There are likely several observational biases driving the numbers of transients of each type being discovered and classified. For example, most optical TDEs have longer rise-times and more luminous peak magnitudes compared to typical Type Ia SNe \citep[e.g.][and references therein]{Maguire2016,vanVelzen2020}, making TDEs easier to discover and observe spectroscopically compared to Type Ia SNe. Both TDEs and Type Ia SNe are more luminous than typical core collapse SNe \citep[e.g.][and references therein]{Arcavi2016,Pian2016}, suppressing the observed fraction of any possible core collapse events in these bright galaxy centers. Also, the properties of the TDEs and their hosts will affect their detectability \citep{Roth2021}.

In addition, here we combine data from various transient surveys and classification campaigns, some focused on classifying the most likely TDE candidates, some possibly looking for Type Ia SNe, and some possibly avoiding likely AGN. This introduces even more complex (and possibly competing) selection effects into the sample of transients. Therefore it is highly nontrivial, if not impossible, to translate these fractions into intrinsic rates \citep[but see][]{Roth2021}. These observed fractions do, however, reflect the current prospects of community classification results when following up discoveries in galaxy centers.  

Naively, spectroscopically identified PS galaxies would thus be the best galaxies to focus a TDE search on, since a transient in such a galaxy is roughly 16 times more likely to be a TDE than a Type Ia SN than if it were in a random galaxy in the \citetalias{French2018} catalog (and about 37 times more likely than if it were in a random quiescent galaxy). Unfortunately, there are only 1683 spectroscopically confirmed PS galaxies in the \citetalias{French2018} catalog, constituting just 2\% of it (middle panel of Figure \ref{fig:no_of_events}). Hence, in absolute numbers, searching for transients in the full catalog will provide roughly 3 times more TDEs, but at the price of having to classify $\sim$8 Type Ia SNe per confirmed TDE (bottom panel of Figure \ref{fig:no_of_events}). Of course, one can also employ photometric classification criteria to newly discovered transients in order to try to reduce Type Ia SN contamination before obtaining spectra.

\subsection{ZTF Search: Light Curves of Unclassified Events}

An important parameter in trying to determine whether an unclassified transient could have been a TDE is the absolute magnitude of its light curve. We search SDSS DR16 and the 2dF Galaxy Redshift Survey \citep[2dFGRS;][]{Colless2003} for host galaxy redshifts of the ZTF photometrically selected events coincident with the center of a galaxy in the \citetalias{French2018} catalog. Our findings are presented in Table \ref{tab:ztf}. For ZTF20abxphdt, a ``Gold'' event, we obtained our own spectrum of the host galaxy and measured the redshift to be 0.0675 from narrow \ion{Ca}{2} H+K and \ion{Na}{1} D absorption features (Fig. \ref{fig:spec})\footnote{Our spectrum was obtained with the Floyds spectrograph mounted on the Las Cumbres Observatory 2-meter telescope in Haleakala, Hawaii \citep{Brown2013}, and was reduced using the \texttt{floydsspec} custom pipeline, which performs flux and wavelength calibration, cosmic-ray removal, and spectrum extraction. The pipeline is available at \url{https://github.com/svalenti/FLOYDS_pipeline/blob/master/ bin/floydsspec/}.}. For each event with a determined redshift, we include an absolute magnitude scale in its light curve in Figures \ref{fig:gold}, \ref{fig:silver} and \ref{fig:bronze}.

The ``Silver'' and ``Bronze'' light curves do not have TDE-like transient behavior \citep[per definition, these are events with no clear rise and decline as seen in optical TDEs;][]{vanVelzen2020}. To determine whether any of the ``Gold'' events might have been a missed TDE, we compare in Figure \ref{fig:gold} each the light curves to those of the prototypical optical TDE PS1-10jh \citep{Gezari2012} and the faint rapidly evolving optical TDE iPTF16fnl \citep{Blagorodnova2017,Brown2018}, whose light curves we obtain from the Open TDE Catalog\footnote{\url{https://tde.space/}}. These two events roughly span the range of known optical TDE light-curve luminosities and time scales (see \citealt{vanVelzen2020} for a review). 

While all of the ``Gold'' light curves have peak absolute luminosities and time scales in the correct range, their $g$--$r$ colors are much redder than those of TDEs. We conclude that none of these events are likely missed TDEs, but transients of some other nature.

\section{Summary and Conclusions}

We quantify the chances of a transient discovered in the center of a galaxy from a catalog of likely TDE hosts to be a TDE or a Type Ia SN (no other types of true transients were classified in these hosts) by searching the classifications of all transients discovered in the centers of these galaxies. The catalog is made up of galaxies selected in different ways, with the bulk being photometrically selected. 

The catalog reduces the contamination of Type Ia SNe by a factor of roughly 2.7 compared to a control sample of quiescent galaxies. The lowest contamination of Type Ia SNe exists in the spectroscopically identified PS subcatalog, but it constitute only 2\% of the entire catalog. By classifying transients from the entire catalog, three times more TDEs are expected to be found, but with a roughly 16 times larger Type Ia SN contamination. 

We have not identified any transients coincident with the center of a galaxy in the catalog as likely missed TDEs.

~\\
We thank O. Yaron for assistance in obtaining TNS data, C. Pellegrino for reducing the ZTF20abxphdt host galaxy spectrum, and M. Nicholl for implementing the \citetalias{French2018} galaxy catalog as a watchlist on Lasair.
I.A. is a CIFAR Azrieli Global Scholar in the Gravity and the Extreme Universe Program and acknowledges support from that program, from the European Research Council (ERC) under the European Union’s Horizon 2020 research and innovation program (grant agreement number 852097), from the Israel Science Foundation (grant number 2752/19), from the United States -- Israel Binational Science Foundation (BSF), and from the Israeli Council for Higher Education Alon Fellowship.
I.N. acknowledges funding from the MIT International Science and Technology Initiatives (MISTI) Israel Program. 

\bibliography{refs}{}

\begin{thebibliography}{}
\expandafter\ifx\csname natexlab\endcsname\relax\def\natexlab#1{#1}\fi
\providecommand{\url}[1]{\href{#1}{#1}}
\providecommand{\dodoi}[1]{doi:~\href{http://doi.org/#1}{\nolinkurl{#1}}}
\providecommand{\doeprint}[1]{\href{http://ascl.net/#1}{\nolinkurl{http://ascl.net/#1}}}
\providecommand{\doarXiv}[1]{\href{https://arxiv.org/abs/#1}{\nolinkurl{https://arxiv.org/abs/#1}}}

\bibitem[{{Abbott} {et~al.}(2018){Abbott}, {Abdalla}, {Allam}, \&
  et~al.}]{Abbott2018}
{Abbott}, T.~M.~C., {Abdalla}, F.~B., {Allam}, S., \& et~al. 2018, \apjs, 239,
  18, \dodoi{10.3847/1538-4365/aae9f0}

\bibitem[{{Alam} {et~al.}(2015){Alam}, {Albareti}, {Allende Prieto}, \&
  et~al.}]{Alam2015}
{Alam}, S., {Albareti}, F.~D., {Allende Prieto}, C., \& et~al. 2015, \apjs,
  219, 12, \dodoi{10.1088/0067-0049/219/1/12}

\bibitem[{Arcavi(2016)}]{Arcavi2016}
Arcavi, I. 2016, Hydrogen-Rich Core-Collapse Supernovae, ed. A.~W. Alsabti \&
  P.~Murdin (Cham: Springer International Publishing), 1--38,
  \dodoi{10.1007/978-3-319-20794-0_39-1}

\bibitem[{{Arcavi} {et~al.}(2014){Arcavi}, {Gal-Yam}, {Sullivan}, {Pan},
  {Cenko}, {Horesh}, {Ofek}, {De Cia}, {Yan}, {Yang}, {Howell}, {Tal},
  {Kulkarni}, {Tendulkar}, {Tang}, {Xu}, {Sternberg}, {Cohen}, {Bloom},
  {Nugent}, {Kasliwal}, {Perley}, {Quimby}, {Miller}, {Theissen}, \&
  {Laher}}]{Arcavi2014}
{Arcavi}, I., {Gal-Yam}, A., {Sullivan}, M., {et~al.} 2014, \apj, 793, 38,
  \dodoi{10.1088/0004-637X/793/1/38}

\bibitem[{{Bellm}(2014)}]{Bellm2014}
{Bellm}, E. 2014, in The Third Hot-wiring the Transient Universe Workshop, ed.
  P.~R. {Wozniak}, M.~J. {Graham}, A.~A. {Mahabal}, \& R.~{Seaman}, 27--33.
\newblock \doarXiv{1410.8185}

\bibitem[{{Blagorodnova} {et~al.}(2017){Blagorodnova}, {Gezari}, {Hung},
  {Kulkarni}, {Cenko}, {Pasham}, {Yan}, {Arcavi}, {Ben-Ami}, {Bue}, {Cantwell},
  {Cao}, {Castro-Tirado}, {Fender}, {Fremling}, {Gal-Yam}, {Ho}, {Horesh},
  {Hosseinzadeh}, {Kasliwal}, {Kong}, {Laher}, {Leloudas}, {Lunnan}, {Masci},
  {Mooley}, {Neill}, {Nugent}, {Powell}, {Valeev}, {Vreeswijk}, {Walters}, \&
  {Wozniak}}]{Blagorodnova2017}
{Blagorodnova}, N., {Gezari}, S., {Hung}, T., {et~al.} 2017, \apj, 844, 46,
  \dodoi{10.3847/1538-4357/aa7579}

\bibitem[{{Brown} {et~al.}(2018){Brown}, {Kochanek}, {Holoien}, {Stanek},
  {Auchettl}, {Shappee}, {Prieto}, {Morrell}, {Falco}, {Strader}, {Chomiuk},
  {Post}, {Villanueva}, {Mathur}, {Dong}, {Chen}, \& {Bose}}]{Brown2018}
{Brown}, J.~S., {Kochanek}, C.~S., {Holoien}, T.~W.~S., {et~al.} 2018, \mnras,
  473, 1130, \dodoi{10.1093/mnras/stx2372}

\bibitem[{{Brown} {et~al.}(2013){Brown}, {Baliber}, {Bianco}, {Bowman},
  {Burleson}, {Conway}, {Crellin}, {Depagne}, {De Vera}, {Dilday}, {Dragomir},
  {Dubberley}, {Eastman}, {Elphick}, {Falarski}, {Foale}, {Ford}, {Fulton},
  {Garza}, {Gomez}, {Graham}, {Greene}, {Haldeman}, {Hawkins}, {Haworth},
  {Haynes}, {Hidas}, {Hjelstrom}, {Howell}, {Hygelund}, {Lister}, {Lobdill},
  {Martinez}, {Mullins}, {Norbury}, {Parrent}, {Paulson}, {Petry}, {Pickles},
  {Posner}, {Rosing}, {Ross}, {Sand}, {Saunders}, {Shobbrook}, {Shporer},
  {Street}, {Thomas}, {Tsapras}, {Tufts}, {Valenti}, {Vander Horst}, {Walker},
  {White}, \& {Willis}}]{Brown2013}
{Brown}, T.~M., {Baliber}, N., {Bianco}, F.~B., {et~al.} 2013, \pasp, 125,
  1031, \dodoi{10.1086/673168}

\bibitem[{{Chambers} {et~al.}(2016){Chambers}, {Magnier}, {Metcalfe},
  {Flewelling}, {Huber}, {Waters}, {Denneau}, {Draper}, {Farrow}, {Finkbeiner},
  {Holmberg}, {Koppenhoefer}, {Price}, {Rest}, {Saglia}, {Schlafly}, {Smartt},
  {Sweeney}, {Wainscoat}, {Burgett}, {Chastel}, {Grav}, {Heasley}, {Hodapp},
  {Jedicke}, {Kaiser}, {Kudritzki}, {Luppino}, {Lupton}, {Monet}, {Morgan},
  {Onaka}, {Shiao}, {Stubbs}, {Tonry}, {White}, {Ba{\~n}ados}, {Bell},
  {Bender}, {Bernard}, {Boegner}, {Boffi}, {Botticella}, {Calamida},
  {Casertano}, {Chen}, {Chen}, {Cole}, {Deacon}, {Frenk}, {Fitzsimmons},
  {Gezari}, {Gibbs}, {Goessl}, {Goggia}, {Gourgue}, {Goldman}, {Grant},
  {Grebel}, {Hambly}, {Hasinger}, {Heavens}, {Heckman}, {Henderson}, {Henning},
  {Holman}, {Hopp}, {Ip}, {Isani}, {Jackson}, {Keyes}, {Koekemoer}, {Kotak},
  {Le}, {Liska}, {Long}, {Lucey}, {Liu}, {Martin}, {Masci}, {McLean}, {Mindel},
  {Misra}, {Morganson}, {Murphy}, {Obaika}, {Narayan}, {Nieto-Santisteban},
  {Norberg}, {Peacock}, {Pier}, {Postman}, {Primak}, {Rae}, {Rai}, {Riess},
  {Riffeser}, {Rix}, {R{\"o}ser}, {Russel}, {Rutz}, {Schilbach}, {Schultz},
  {Scolnic}, {Strolger}, {Szalay}, {Seitz}, {Small}, {Smith}, {Soderblom},
  {Taylor}, {Thomson}, {Taylor}, {Thakar}, {Thiel}, {Thilker}, {Unger},
  {Urata}, {Valenti}, {Wagner}, {Walder}, {Walter}, {Watters}, {Werner},
  {Wood-Vasey}, \& {Wyse}}]{Chambers2016}
{Chambers}, K.~C., {Magnier}, E.~A., {Metcalfe}, N., {et~al.} 2016, arXiv
  e-prints, arXiv:1612.05560.
\newblock \doarXiv{1612.05560}

\bibitem[{Clopper \& Pearson(1934)}]{clopper1934}
Clopper, C.~J., \& Pearson, E.~S. 1934, Biometrika, 26, 404,
  \dodoi{10.1093/biomet/26.4.404}

\bibitem[{{Colless} {et~al.}(2003){Colless}, {Peterson}, {Jackson}, {Peacock},
  {Cole}, {Norberg}, {Baldry}, {Baugh}, {Bland-Hawthorn}, {Bridges}, {Cannon},
  {Collins}, {Couch}, {Cross}, {Dalton}, {De Propris}, {Driver}, {Efstathiou},
  {Ellis}, {Frenk}, {Glazebrook}, {Lahav}, {Lewis}, {Lumsden}, {Maddox},
  {Madgwick}, {Sutherland}, \& {Taylor}}]{Colless2003}
{Colless}, M., {Peterson}, B.~A., {Jackson}, C., {et~al.} 2003, arXiv e-prints,
  astro.
\newblock \doarXiv{astro-ph/0306581}

\bibitem[{{F{\"o}rster} {et~al.}(2021){F{\"o}rster}, {Cabrera-Vives},
  {Castillo-Navarrete}, {Est{\'e}vez}, {S{\'a}nchez-S{\'a}ez}, {Arredondo},
  {Bauer}, {Carrasco-Davis}, {Catelan}, {Elorrieta}, {Eyheramendy}, {Huijse},
  {Pignata}, {Reyes}, {Reyes}, {Rodr{\'\i}guez-Mancini}, {Ruz-Mieres},
  {Valenzuela}, {{\'A}lvarez-Maldonado}, {Astorga}, {Borissova}, {Clocchiatti},
  {De Cicco}, {Donoso-Oliva}, {Hern{\'a}ndez-Garc{\'\i}a}, {Graham},
  {Jord{\'a}n}, {Kurtev}, {Mahabal}, {Maureira}, {Mu{\~n}oz-Arancibia},
  {Molina-Ferreiro}, {Moya}, {Palma}, {P{\'e}rez-Carrasco}, {Protopapas},
  {Romero}, {Sabatini-Gacitua}, {S{\'a}nchez}, {San Mart{\'\i}n},
  {Sep{\'u}lveda-Cobo}, {Vera}, \& {Vergara}}]{Forster2021}
{F{\"o}rster}, F., {Cabrera-Vives}, G., {Castillo-Navarrete}, E., {et~al.}
  2021, \aj, 161, 242, \dodoi{10.3847/1538-3881/abe9bc}

\bibitem[{{French} {et~al.}(2016){French}, {Arcavi}, \&
  {Zabludoff}}]{French2016}
{French}, K.~D., {Arcavi}, I., \& {Zabludoff}, A. 2016, \apjl, 818, L21,
  \dodoi{10.3847/2041-8205/818/1/L21}

\bibitem[{{French} {et~al.}(2020){French}, {Wevers}, {Law-Smith}, {Graur}, \&
  {Zabludoff}}]{French2020}
{French}, K.~D., {Wevers}, T., {Law-Smith}, J., {Graur}, O., \& {Zabludoff},
  A.~I. 2020, \ssr, 216, 32, \dodoi{10.1007/s11214-020-00657-y}

\bibitem[{{French} \& {Zabludoff}(2018)}]{French2018}
{French}, K.~D., \& {Zabludoff}, A.~I. 2018, \apj, 868, 99,
  \dodoi{10.3847/1538-4357/aaea64}

\bibitem[{{Gehrels}(1986)}]{Gehrels1986}
{Gehrels}, N. 1986, \apj, 303, 336, \dodoi{10.1086/164079}

\bibitem[{{Gezari} {et~al.}(2012){Gezari}, {Chornock}, {Rest}, {Huber},
  {Forster}, {Berger}, {Challis}, {Neill}, {Martin}, {Heckman}, {Lawrence},
  {Norman}, {Narayan}, {Foley}, {Marion}, {Scolnic}, {Chomiuk}, {Soderberg},
  {Smith}, {Kirshner}, {Riess}, {Smartt}, {Stubbs}, {Tonry}, {Wood-Vasey},
  {Burgett}, {Chambers}, {Grav}, {Heasley}, {Kaiser}, {Kudritzki}, {Magnier},
  {Morgan}, \& {Price}}]{Gezari2012}
{Gezari}, S., {Chornock}, R., {Rest}, A., {et~al.} 2012, \nat, 485, 217,
  \dodoi{10.1038/nature10990}

\bibitem[{{Graham} {et~al.}(2019){Graham}, {Kulkarni}, {Bellm}, {Adams},
  {Barbarino}, {Blagorodnova}, {Bodewits}, {Bolin}, {Brady}, {Cenko}, {Chang},
  {Coughlin}, {De}, {Eadie}, {Farnham}, {Feindt}, {Franckowiak}, {Fremling},
  {Gezari}, {Ghosh}, {Goldstein}, {Golkhou}, {Goobar}, {Ho}, {Huppenkothen},
  {Ivezi{\'c}}, {Jones}, {Juric}, {Kaplan}, {Kasliwal}, {Kelley}, {Kupfer},
  {Lee}, {Lin}, {Lunnan}, {Mahabal}, {Miller}, {Ngeow}, {Nugent}, {Ofek},
  {Prince}, {Rauch}, {van Roestel}, {Schulze}, {Singer}, {Sollerman}, {Taddia},
  {Yan}, {Ye}, {Yu}, {Barlow}, {Bauer}, {Beck}, {Belicki}, {Biswas}, {Brinnel},
  {Brooke}, {Bue}, {Bulla}, {Burruss}, {Connolly}, {Cromer}, {Cunningham},
  {Dekany}, {Delacroix}, {Desai}, {Duev}, {Feeney}, {Flynn}, {Frederick},
  {Gal-Yam}, {Giomi}, {Groom}, {Hacopians}, {Hale}, {Helou}, {Henning},
  {Hover}, {Hillenbrand}, {Howell}, {Hung}, {Imel}, {Ip}, {Jackson}, {Kaspi},
  {Kaye}, {Kowalski}, {Kramer}, {Kuhn}, {Landry}, {Laher}, {Mao}, {Masci},
  {Monkewitz}, {Murphy}, {Nordin}, {Patterson}, {Penprase}, {Porter},
  {Rebbapragada}, {Reiley}, {Riddle}, {Rigault}, {Rodriguez}, {Rusholme}, {van
  Santen}, {Shupe}, {Smith}, {Soumagnac}, {Stein}, {Surace}, {Szkody}, {Terek},
  {Van Sistine}, {van Velzen}, {Vestrand}, {Walters}, {Ward}, {Zhang}, \&
  {Zolkower}}]{Graham2019}
{Graham}, M.~J., {Kulkarni}, S.~R., {Bellm}, E.~C., {et~al.} 2019, \pasp, 131,
  078001, \dodoi{10.1088/1538-3873/ab006c}

\bibitem[{{Hinshaw} {et~al.}(2013){Hinshaw}, {Larson}, {Komatsu}, {Spergel},
  {Bennett}, {Dunkley}, {Nolta}, {Halpern}, {Hill}, {Odegard}, {Page}, {Smith},
  {Weiland}, {Gold}, {Jarosik}, {Kogut}, {Limon}, {Meyer}, {Tucker}, {Wollack},
  \& {Wright}}]{Hinshaw2013}
{Hinshaw}, G., {Larson}, D., {Komatsu}, E., {et~al.} 2013, \apjs, 208, 19,
  \dodoi{10.1088/0067-0049/208/2/19}

\bibitem[{Maguire(2016)}]{Maguire2016}
Maguire, K. 2016, Type Ia Supernovae, ed. A.~W. Alsabti \& P.~Murdin (Cham:
  Springer International Publishing), 1--24,
  \dodoi{10.1007/978-3-319-20794-0_36-1}

\bibitem[{Pian \& Mazzali(2016)}]{Pian2016}
Pian, E., \& Mazzali, P.~A. 2016, Hydrogen-Poor Core-Collapse Supernovae, ed.
  A.~W. Alsabti \& P.~Murdin (Cham: Springer International Publishing), 1--16,
  \dodoi{10.1007/978-3-319-20794-0_40-1}

\bibitem[{{Rees}(1988)}]{Rees1988}
{Rees}, M.~J. 1988, \nat, 333, 523, \dodoi{10.1038/333523a0}

\bibitem[{{Roth} {et~al.}(2021){Roth}, {van Velzen}, {Cenko}, \&
  {Mushotzky}}]{Roth2021}
{Roth}, N., {van Velzen}, S., {Cenko}, S.~B., \& {Mushotzky}, R.~F. 2021, \apj,
  910, 93, \dodoi{10.3847/1538-4357/abdf50}

\bibitem[{{Smith} {et~al.}(2019){Smith}, {Williams}, {Young}, {Ibsen},
  {Smartt}, {Lawrence}, {Morris}, {Voutsinas}, \& {Nicholl}}]{Smith2019}
{Smith}, K.~W., {Williams}, R.~D., {Young}, D.~R., {et~al.} 2019, Research
  Notes of the American Astronomical Society, 3, 26,
  \dodoi{10.3847/2515-5172/ab020f}

\bibitem[{{Smith} {et~al.}(2020){Smith}, {Smartt}, {Young}, {Tonry}, {Denneau},
  {Flewelling}, {Heinze}, {Weiland}, {Stalder}, {Rest}, {Stubbs}, {Anderson},
  {Chen}, {Clark}, {Do}, {F{\"o}rster}, {Fulton}, {Gillanders}, {McBrien},
  {O'Neill}, {Srivastav}, \& {Wright}}]{Smith2020}
{Smith}, K.~W., {Smartt}, S.~J., {Young}, D.~R., {et~al.} 2020, \pasp, 132,
  085002, \dodoi{10.1088/1538-3873/ab936e}

\bibitem[{{Stone} \& {Metzger}(2016)}]{Stone2016}
{Stone}, N.~C., \& {Metzger}, B.~D. 2016, \mnras, 455, 859,
  \dodoi{10.1093/mnras/stv2281}

\bibitem[{{Strauss} {et~al.}(2002){Strauss}, {Weinberg}, {Lupton}, {Narayanan},
  {Annis}, {Bernardi}, {Blanton}, {Burles}, {Connolly}, {Dalcanton}, {Doi},
  {Eisenstein}, {Frieman}, {Fukugita}, {Gunn}, {Ivezi{\'c}}, {Kent}, {Kim},
  {Knapp}, {Kron}, {Munn}, {Newberg}, {Nichol}, {Okamura}, {Quinn}, {Richmond},
  {Schlegel}, {Shimasaku}, {SubbaRao}, {Szalay}, {Vanden Berk}, {Vogeley},
  {Yanny}, {Yasuda}, {York}, \& {Zehavi}}]{Strauss2002}
{Strauss}, M.~A., {Weinberg}, D.~H., {Lupton}, R.~H., {et~al.} 2002, \aj, 124,
  1810, \dodoi{10.1086/342343}

\bibitem[{{van Velzen} {et~al.}(2020){van Velzen}, {Holoien}, {Onori}, {Hung},
  \& {Arcavi}}]{vanVelzen2020}
{van Velzen}, S., {Holoien}, T. W.~S., {Onori}, F., {Hung}, T., \& {Arcavi}, I.
  2020, \ssr, 216, 124, \dodoi{10.1007/s11214-020-00753-z}

\bibitem[{{Wang} \& {Merritt}(2004)}]{Wang2004}
{Wang}, J., \& {Merritt}, D. 2004, \apj, 600, 149, \dodoi{10.1086/379767}

\bibitem[{{Wright} {et~al.}(2010){Wright}, {Eisenhardt}, {Mainzer}, {Ressler},
  {Cutri}, {Jarrett}, {Kirkpatrick}, {Padgett}, {McMillan}, {Skrutskie},
  {Stanford}, {Cohen}, {Walker}, {Mather}, {Leisawitz}, {Gautier}, {McLean},
  {Benford}, {Lonsdale}, {Blain}, {Mendez}, {Irace}, {Duval}, {Liu}, {Royer},
  {Heinrichsen}, {Howard}, {Shannon}, {Kendall}, {Walsh}, {Larsen}, {Cardon},
  {Schick}, {Schwalm}, {Abid}, {Fabinsky}, {Naes}, \& {Tsai}}]{Wright2010}
{Wright}, E.~L., {Eisenhardt}, P. R.~M., {Mainzer}, A.~K., {et~al.} 2010, \aj,
  140, 1868, \dodoi{10.1088/0004-6256/140/6/1868}

\bibitem[{{York} {et~al.}(2000){York}, {Adelman}, {Anderson}, {Anderson},
  {Annis}, {Bahcall}, {Bakken}, {Barkhouser}, {Bastian}, {Berman}, {Boroski},
  {Bracker}, {Briegel}, {Briggs}, {Brinkmann}, {Brunner}, {Burles}, {Carey},
  {Carr}, {Castander}, {Chen}, {Colestock}, {Connolly}, {Crocker}, {Csabai},
  {Czarapata}, {Davis}, {Doi}, {Dombeck}, {Eisenstein}, {Ellman}, {Elms},
  {Evans}, {Fan}, {Federwitz}, {Fiscelli}, {Friedman}, {Frieman}, {Fukugita},
  {Gillespie}, {Gunn}, {Gurbani}, {de Haas}, {Haldeman}, {Harris}, {Hayes},
  {Heckman}, {Hennessy}, {Hindsley}, {Holm}, {Holmgren}, {Huang}, {Hull},
  {Husby}, {Ichikawa}, {Ichikawa}, {Ivezi{\'c}}, {Kent}, {Kim}, {Kinney},
  {Klaene}, {Kleinman}, {Kleinman}, {Knapp}, {Korienek}, {Kron}, {Kunszt},
  {Lamb}, {Lee}, {Leger}, {Limmongkol}, {Lindenmeyer}, {Long}, {Loomis},
  {Loveday}, {Lucinio}, {Lupton}, {MacKinnon}, {Mannery}, {Mantsch}, {Margon},
  {McGehee}, {McKay}, {Meiksin}, {Merelli}, {Monet}, {Munn}, {Narayanan},
  {Nash}, {Neilsen}, {Neswold}, {Newberg}, {Nichol}, {Nicinski}, {Nonino},
  {Okada}, {Okamura}, {Ostriker}, {Owen}, {Pauls}, {Peoples}, {Peterson},
  {Petravick}, {Pier}, {Pope}, {Pordes}, {Prosapio}, {Rechenmacher}, {Quinn},
  {Richards}, {Richmond}, {Rivetta}, {Rockosi}, {Ruthmansdorfer}, {Sandford},
  {Schlegel}, {Schneider}, {Sekiguchi}, {Sergey}, {Shimasaku}, {Siegmund},
  {Smee}, {Smith}, {Snedden}, {Stone}, {Stoughton}, {Strauss}, {Stubbs},
  {SubbaRao}, {Szalay}, {Szapudi}, {Szokoly}, {Thakar}, {Tremonti}, {Tucker},
  {Uomoto}, {Vanden Berk}, {Vogeley}, {Waddell}, {Wang}, {Watanabe},
  {Weinberg}, {Yanny}, {Yasuda}, \& {SDSS Collaboration}}]{York2000}
{York}, D.~G., {Adelman}, J., {Anderson}, John~E., J., {et~al.} 2000, \aj, 120,
  1579, \dodoi{10.1086/301513}

\end{thebibliography}
\bibliographystyle{aasjournal}

\appendix

\section{TNS Events}

We list in Table \ref{tab:tns} the full set of TNS events coincident with the center of a \citetalias{French2018} galaxy. For each event we note, among other information, its type, if it is spectroscopically classified on the TNS, its additional survey names, if they were reported to the TNS, and its separation from the host center, according to the TNS coordinates of the event and host coordinates in \citetalias{French2018}.

\begin{longtable}{lDDlp{0.13\linewidth}ll}
\caption{\label{tab:tns}Events in the TNS Coincident within 1\arcsec\ with the Center of a Galaxy in the \citetalias{French2018} Catalog.}\\
\hline
IAU Name & \multicolumn2c{RA} & \multicolumn2c{Dec} & Classification & Other Name(s) & FZ18 Table(s) & Separation From Host \\
 & \multicolumn2c{[deg]} & \multicolumn2c{[deg]} & & & & [\arcsec] \\
\hline
\hline
\decimals
SN 1999du & 16.77475 & -0.13161 & SN Ia &  & 5 & 0.95 \\
SN 2001G & 137.38825 & 50.28092 & SN Ia &  & 1 & 0.80 \\
SN 2006ae & 222.09696 & 21.79764 & SN Ia &  & 5 & 0.15 \\
SN 2007mj & 53.68517 & 0.35553 & SN Ia &  & 1 & 0.80 \\
SN 2009fx & 253.29700 & 23.96525 & SN Ia &  & 1 & 0.43 \\
SN 2016aud & 228.47623 & 4.75726 & SN Ia &  & 1,2 & 0.43 \\
AT 2016sq & 195.87621 & 19.77461 & Not Classified & PS16ug & 1,2 & 0.83 \\
AT 2017akd & 161.02167 & 20.57111 & Not Classified & PS17aql & 5 & 0.79 \\
AT 2017bcg & 211.52899 & 42.67363 & Not Classified & iPTF17bcg & 1 & 0.76 \\
AT 2017br & 128.71599 & 44.24706 & Not Classified & iPTF17br & 1 & 0.06 \\
AT 2017brz & 152.29958 & 1.63526 & Not Classified & iPTF17brz & 1 & 0.68 \\
AT 2017bs & 128.82731 & 44.02291 & Not Classified &  & 1 & 0.18 \\
AT 2017cks & 162.49356 & 38.09399 & Not Classified &  & 1 & 0.08 \\
AT 2017eu & 142.32666 & 62.35766 & Not Classified &  & 1 & 0.68 \\
AT 2017hvp & 178.31441 & 15.06375 & Not Classified &  & 1 & 0.55 \\
AT 2018ahh & 161.20269 & 51.95058 & Not Classified & ATLAS18mlv, PS18amw & 1 & 0.39 \\
AT 2018ail & 196.08738 & 1.15093 & Galaxy &  & 1 & 0.26 \\
AT 2018bpb & 213.65545 & 61.12649 & Not Classified & ATLAS18opw & 5 & 0.33 \\
AT 2018bvy & 180.51065 & 15.01067 & Not Classified & ZTF18aastwrz & 1 & 0.18 \\
AT 2018cqh & 38.44554 & -1.02455 & Not Classified &  & 1 & 0.08 \\
SN 2018ffi & 325.04707 & 21.55839 & SN Ia &  & 5 & 0.69 \\
AT 2018fyv & 244.06311 & 21.00398 & Not Classified & PS18bnd & 5 & 0.11 \\
SN 2018gvb & 249.64330 & 31.87302 & SN Ia & ZTF18abwmuua, ATLAS18vsb & 1 & 0.42 \\
AT 2018hym & 228.45355 & 45.82884 & Not Classified & ATLAS18sjn & 1 & 0.19 \\
AT 2018hyz & 151.71196 & 1.69280 & TDE & ASASSN-18zj, ZTF18acpdvos, ATLAS18bafs & 1,2 & 0.10 \\
AT 2018jkl & 153.45776 & 39.64185 & Not Classified & ZTF18acsoyiv & 1 & 0.08 \\
AT 2018kmp & 202.17649 & 55.36520 & Not Classified &  & 1 & 0.64 \\
AT 2019aale & 88.35786 & 11.38557 & AGN & & 5 & 0.07 \\
AT 2019azh & 123.32061 & 22.64834 & TDE & ASASSN-19dj, ZTF17aaazdba, Gaia19bvo, ZTF18achzddr & 1,2 & 0.16 \\
AT 2019bvk & 192.01691 & 18.97612 & Not Classified &  & 1 & 0.82 \\
SN 2019bxh & 222.97121 & 55.40762 & SN Ia &  & 1 & 0.86 \\
AT 2019cxz & 161.54031 & 24.26684 & Not Classified & ZTF19aaoxijx, ATLAS19ghr & 1 & 0.22 \\
AT 2019dec & 186.84407 & 30.09247 & Not Classified &  & 1 & 0.26 \\
AT 2019ev & 106.68529 & 18.52649 & Not Classified &  & 5 & 0.69 \\
AT 2019gq & 120.01922 & 58.76122 & Not Classified &  & 5 & 0.25 \\
SN 2019kcj & 242.67834 & 15.15591 & SN Ia &  & 1 & 0.83 \\
AT 2019lwx & 184.82792 & 15.77210 & Not Classified &  & 1 & 0.12 \\
AT 2019mds & 248.50573 & 77.60043 & Not Classified &  & 5 & 0.69 \\
AT 2019nks & 97.10029 & -17.28797 & Not Classified & Gaia19dmk & 5 & 0.13 \\
SN 2019nvh & 237.65586 & 23.04686 & SN Ia & ZTF19abpuikr & 1 & 0.23 \\
SN 2019oc & 120.01929 & 40.07781 & SN Ia &  & 1 & 0.43 \\
AT 2019ofz & 126.61545 & 8.74346 & Not Classified & ZTF18acidntq & 1 & 0.17 \\
AT 2019rru & 117.01817 & 22.80833 & Not Classified & ZTF19acbmotn & 1 & 0.07 \\
AT 2019rvr & 222.22627 & 31.65809 & Not Classified &  & 1,2 & 0.37 \\
SN 2019vtx & 124.39083 & 56.04854 & SN Ia & ATLAS19bbyw, ZTF19acxlcdz & 1 & 0.26 \\
SN 2019vva & 346.50381 & 0.31130 & SN Ia & ATLAS19bcew, ZTF19acxgwid & 1 & 0.41 \\
AT 2019xdn & 325.53655 & -7.33813 & Not Classified & ZTF19adakvxy & 1,2 & 0.32 \\
AT 2019xeg & 148.11258 & 36.71760 & Not Classified & ZTF19aczjwxq & 1 & 0.46 \\
SN 2020K & 147.52712 & 5.72366 & SN Ia & ZTF20aaawbkz & 1 & 0.33 \\
AT 2020aajc & 98.97280 & 36.69855 & Not Classified &  & 5 & 0.64 \\
AT 2020abg & 154.10578 & 42.09326 & Not Classified &  & 1 & 0.46 \\
AT 2020adfy & 136.58307 & 52.36386 & Not Classified &  & 1,2 & 0.13 \\
AT 2020aqw & 52.20150 & -4.21910 & Not Classified &  & 5 & 0.19 \\
AT 2020as & 157.73045 & -19.45978 & Not Classified &  & 5 & 0.45 \\
AT 2020bub & 123.50916 & 21.36741 & Not Classified &  & 1 & 0.72 \\
AT 2020bwk & 229.53838 & 45.17097 & Not Classified &  & 7,10 & 0.56 \\
AT 2020cj & 192.39751 & 47.83122 & Not Classified & ATLAS20ank & 1 & 0.40 \\
AT 2020fwm & 229.41903 & 13.85177 & Not Classified & ZTF20aauolxq & 1,2 & 0.21 \\
AT 2020hpd & 252.51063 & 26.42812 & Not Classified &  & 1 & 0.27 \\
SN 2020ism & 208.28775 & 47.29950 & SN Ia & ZTF20aavynba, ATLAS20lvt & 1 & 0.55 \\
AT 2020iwa & 241.15838 & 26.61666 & Not Classified &  & 1 & 0.79 \\
AT 2020jeu & 156.96197 & 2.60946 & Not Classified & PS20ctp & 1 & 0.54 \\
SN 2020jfn & 138.76949 & 10.25872 & SN Ia &  & 1 & 0.53 \\
AT 2020jgq & 237.06576 & 21.02477 & Not Classified & PS20cuf & 1 & 0.32 \\
SN 2020jny & 180.74371 & 20.08543 & SN Ia &  & 1 & 0.93 \\
AT 2020kxs & 331.24181 & 1.00131 & Not Classified &  & 1 & 0.13 \\
AT 2020lce & 313.23469 & 3.42653 & Not Classified & PS20dki & 5 & 0.97 \\
AT 2020nmi & 191.82891 & 44.45004 & Not Classified &  & 1 & 0.54 \\
SN 2020rba & 23.97291 & 39.95597 & SN Ia &  & 5 & 0.13 \\
AT 2020skf & 39.25393 & 26.62204 & Not Classified &  & 5 & 0.10 \\
AT 2020vao & 252.52531 & 32.30183 & Not Classified & ATLAS20bcop, ZTF20abimsxj & 1 & 0.29 \\
SN 2020vf & 182.90418 & 0.32232 & SN Ia & ZTF20aafcjln, ATLAS20auq, PS20qz & 1 & 0.89 \\
AT 2020vgn & 231.77848 & 34.06877 & Not Classified & ZTF19aanleed & 1 & 0.24 \\
AT 2020wey & 136.35783 & 61.80255 & TDE &  & 1 & 0.11 \\
AT 2020xna & 93.25584 & 70.34574 & Not Classified & ZTF20aclflri & 5 & 0.99 \\
SN 2020xsr & 142.20817 & 46.65334 & SN Ia &  & 1 & 0.07 \\
AT 2020ybp & 244.22378 & 31.31074 & Not Classified & ZTF18aavkrsj & 1 & 0.81 \\
AT 2020ygl & 138.30031 & 5.13456 & Not Classified &  & 1 & 0.23 \\
AT 2020yln & 30.90734 & 0.35947 & Not Classified & PS20kme & 1,2 & 0.85 \\
AT 2020zet & 155.16050 & -3.72955 & Not Classified & Gaia20fdo & 5 & 0.07 \\
AT 2021abe & 79.57412 & 58.44900 & Not Classified &  & 5 & 0.34 \\
AT 2021bwg & 225.31005 & 13.02212 & Not Classified & ZTF21aahfizx, ATLAS21egb & 1 & 0.76 \\
SN 2021cky & 229.39751 & 18.11807 & SN Ia & ATLAS21eev, ZTF21aahfjbs & 1 & 0.04 \\
AT 2021duz & 207.86825 & 40.44792 & Not Classified & ZTF18aaviokz & 1 & 0.26 \\
AT 2021ee & 1.42135 & 4.13466 & Not Classified & ZTF21aaaqmuf & 6 & 0.91 \\
SN 2021hmc & 148.62746 & 53.16909 & SN Ia & ZTF18acsremz & 1 & 0.39 \\
AT 2021igj & 223.81404 & 13.65745 & Not Classified & ATLAS21kcw & 1,2 & 0.64 \\
AT 2021ksh & 199.28991 & 59.75154 & Not Classified & ZTF21aawehzm & 1 & 0.90 \\
SN 2021kun & 166.41407 & 19.46029 & SN Ia & ZTF21aaxxjen, ATLAS21ofj & 1 & 0.83 \\
AT 2021kxv & 123.96870 & 29.80600 & Not Classified & ZTF21aaycwpu & 1 & 0.58 \\
AT 2021lkq & 316.71028 & 10.73572 & Not Classified & ZTF21aazmjaf, ATLAS21ovp & 1 & 0.71 \\
AT 2021lml & 215.61896 & 31.75931 & Not Classified & ZTF21aaxyfzb & 1 & 0.67 \\
AT 2021mkd & 233.68061 & 56.99324 & Not Classified & ZTF21abbsncs & 1 & 0.67 \\
AT 2021no & 70.98989 & -38.21690 & Not Classified &  & 6 & 0.83 \\
AT 2021nzg & 216.72033 & 23.16105 & Not Classified & ZTF21abcpqpv & 1 & 0.91 \\
AT 2021osm & 330.12554 & 21.05220 & Not Classified & ZTF21abebkok & 7 & 0.44 \\
SN 2021qus & 196.26836 & 60.77193 & SN Ia & ZTF21abhqqwq & 7,10 & 0.16 \\
AT 2021rw & 78.70566 & 48.03333 & Not Classified &  & 5 & 0.37 \\
AT 2021spt & 240.54084 & 6.64751 & Not Classified & ZTF21abkayhx & 1 & 0.57 \\
AT 2021uhf & 233.22212 & 13.10198 & Not Classified & PS21iil, ATLAS21bdxd & 5 & 0.61 \\
AT 2021vp & 187.14868 & 27.67625 & Not Classified &  & 1 & 0.63 \\
\hline
\end{longtable}

\section{ZTF Events}

We list in Table \ref{tab:ztf} the full set of events in the ZTF public alert stream coincident with an \citetalias{French2018} galaxy, retrieved from the Lasair broker. For each event, we note, among other information, its Sherlock (Lasair machine-learning contextual) classification, its TNS name and spectroscopic classification (if they exist), and its redshift (from the host galaxy spectrum, if available, or from other sources - see main text for details). We also note its light curve ``rank'' if it has sufficient detections (see main text for details).

\begin{figure}[b]
\includegraphics[width=\textwidth]{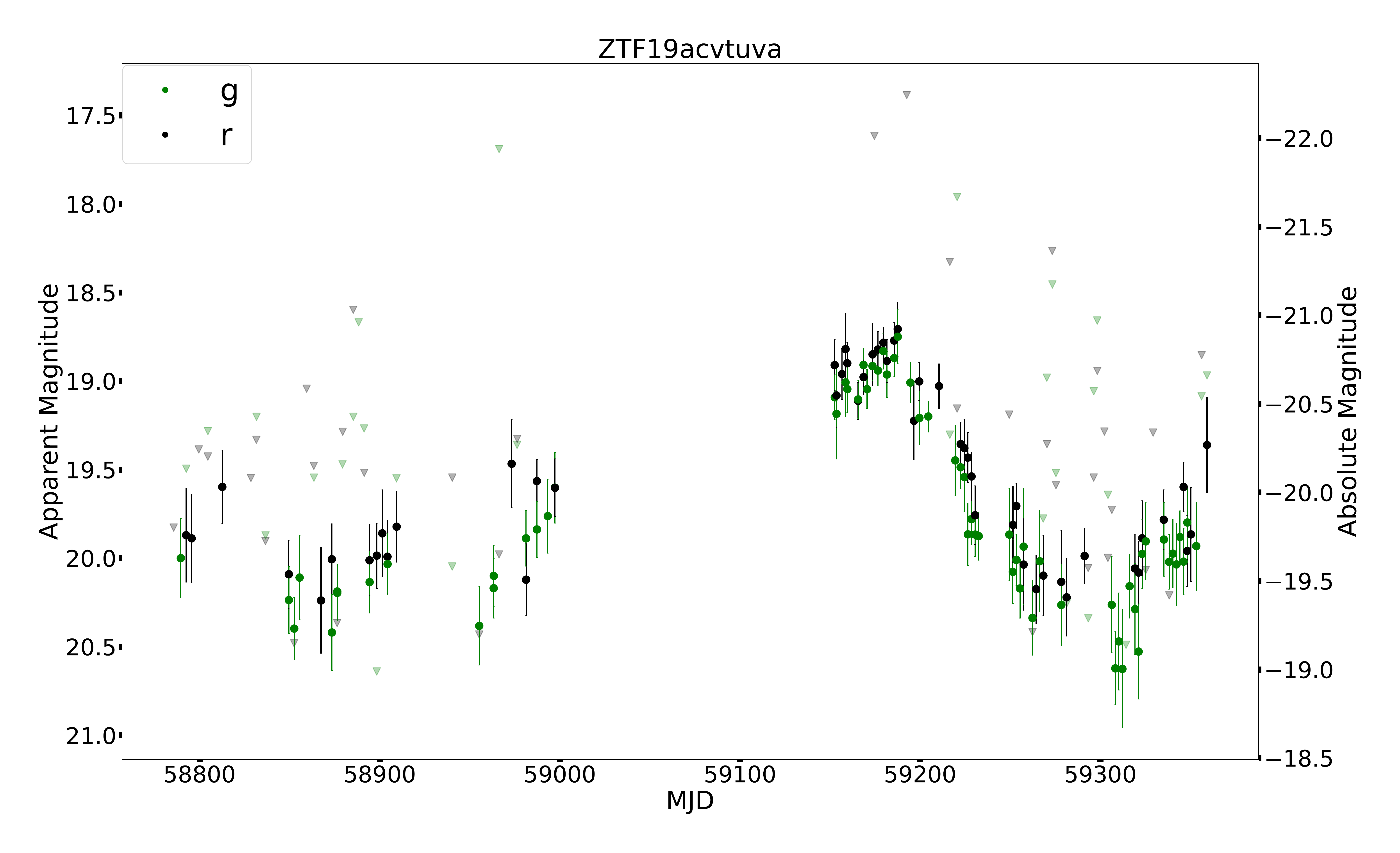}
\caption{\label{fig:silver}ZTF light curve of our ``Silver'' unclassified transient coincident with the center of a galaxy in the \citetalias{French2018} catalog (triangles denote 5$\sigma$ non-detection upper limits). This event appears to be variable rather than transient.} 
\end{figure}

\begin{figure}
\centering
\includegraphics[width=0.9\textwidth]{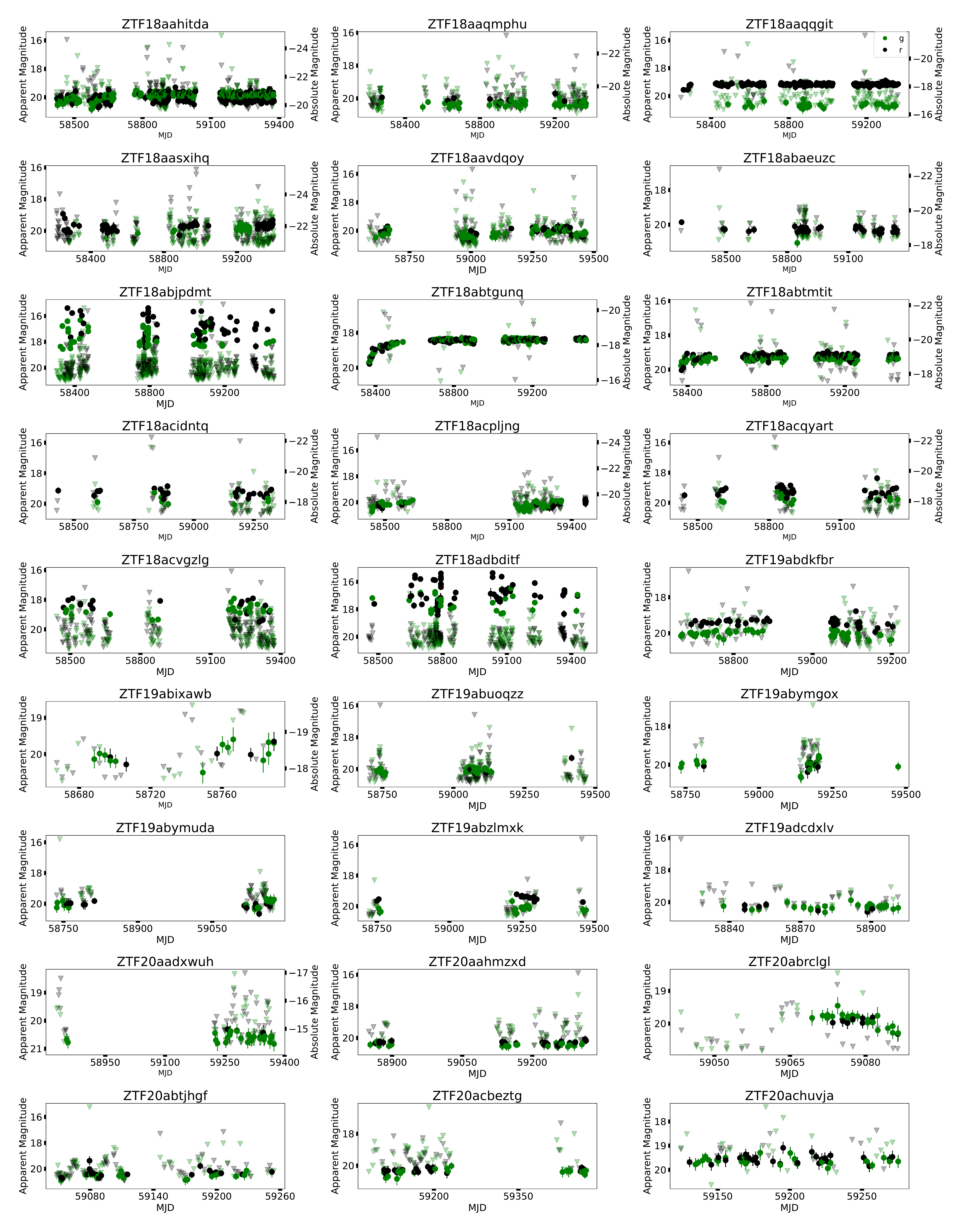}
\caption{\label{fig:bronze}ZTF light curves of our ``Bronze'' set of unclassified transients coincident with the center of a galaxy in the \citetalias{French2018} catalog (triangles denote 5$\sigma$  non-detection upper limits). These events either show a rise followed by constant brightness, or have upper-limits intertwined with the detections, indicating they are either subtraction artifacts or flaring Galactic sources.} 
\end{figure}

\begin{figure}
\includegraphics[width=\textwidth]{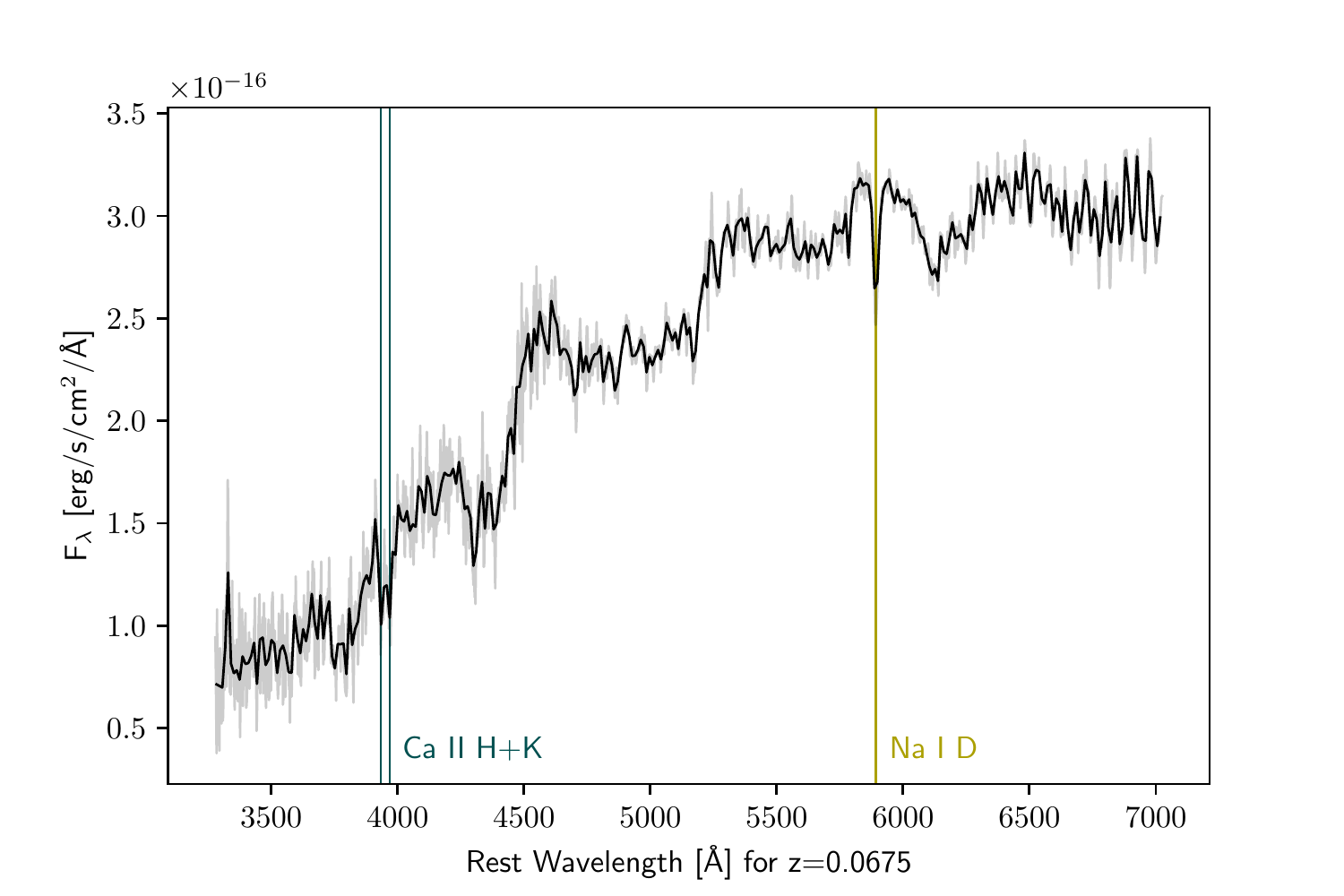}
\caption{\label{fig:spec}Floyds spectrum of the host galaxy of ZTF20abxphdt used to determine its redshift from the \ion{Ca}{2} H+K and \ion{Na}{1} D lines (marked). The original spectrum is shown in gray, and a binned spectrum in black.} 
\end{figure}

We also provide the light curves, as retrieved from the ALeRCE broker, for the ``silver'' (Fig. \ref{fig:silver}) and ``bronze'' (Fig. \ref{fig:bronze}) sets of events, and the spectrum we obtained of the host galaxy of ZTF20abxphdt used to determine its redshift (Fig. \ref{fig:spec}).

\begin{longtable}{lDDlllllll}
\caption{\label{tab:ztf}Events in the ZTF Public Alert Stream Coincident within 1\arcsec\ With the Center of a Galaxy in the \citetalias{French2018} Catalog.}\\

\hline
ZTF Name & \multicolumn2c{RA} & \multicolumn2c{Dec} & No. of & Sherlock & TNS & TNS & FZ18 & Rank & Redshift \\
& \multicolumn2c{} & \multicolumn2c{} & Detections & Class$^a$ & Name & Class & Table(s) & & \\
\hline
\hline
\decimals

ZTF17aaajpbn & 176.80933 & 14.14249 & 4 & NT &  &  & 1 &  &  \\
ZTF17aaazdba & 123.32063 & 22.64830 & 111 & NT & AT 2019azh & TDE & 1,2 &  &  \\
ZTF17aabwnst & 189.25383 & 29.66105 & 2 & NT &  &  & 1 &  &  \\
ZTF17aabxbwj & 204.63031 & 33.17805 & 2 & NT &  &  & 1 &  &  \\
ZTF17aaclxhm & 43.78316 & -3.69214 & 2 & SN &  &  & 6 &  &  \\
ZTF17aacvwvn & 131.40772 & 36.93465 & 3 & NT &  &  & 1 &  &  \\
ZTF18aaajljy & 184.82790 & 15.77213 & 2 & NT & AT 2019lwx &  & 1 &  &  \\
ZTF18aaapdih & 197.89169 & 18.33195 & 3 & NT &  &  & 1 &  &  \\
ZTF18aaapdnx & 177.43238 & 23.44518 & 5 & NT &  &  & 1 &  &  \\
ZTF18aabduzj & 194.59098 & 27.96778 & 4 & AGN &  &  & 5 &  &  \\
ZTF18aabdvaj & 194.61838 & 27.55939 & 2 & NT &  &  & 1,2 &  &  \\
ZTF18aabdvak & 194.32420 & 27.81093 & 2 & NT &  &  & 1 &  &  \\
ZTF18aabdvqf & 183.38759 & 28.86269 & 2 & NT &  &  & 1 &  &  \\
ZTF18aabeibv & 204.80101 & 28.40470 & 4 & NT &  &  & 1,2 &  &  \\
ZTF18aabkjjg & 147.64402 & 44.32343 & 4 & NT &  &  & 1,5 &  &  \\
ZTF18aabvmer & 179.28363 & 26.49880 & 4 & NT &  &  & 1 &  &  \\
ZTF18aabvotu & 187.14849 & 27.67622 & 6 & NT & AT 2021vp &  & 1 &  &  \\
ZTF18aacbnkp & 183.11203 & 29.14921 & 4 & NT &  &  & 1 &  &  \\
ZTF18aacbvcr & 198.16849 & 18.01383 & 2 & NT &  &  & 1 &  &  \\
ZTF18aaccrsz & 199.42068 & 17.69778 & 4 & NT &  &  & 1,2 &  &  \\
ZTF18aadurgi & 98.97258 & 36.69853 & 9 & SN & AT 2020aajc &  & 5 &  &  \\
ZTF18aafomqo & 106.03674 & 21.89812 & 2 & VS &  &  & 5,8 &  &  \\
ZTF18aaggkrv & 154.38140 & 46.98529 & 2 & NT &  &  & 1 &  &  \\
ZTF18aaguaep & 187.01362 & 19.29261 & 2 & NT &  &  & 1,2 &  &  \\
ZTF18aagvaym & 196.24128 & 53.78086 & 10 & NT &  &  & 1 &  &  \\
ZTF18aagydai & 213.87528 & 41.01925 & 2 & VS &  &  & 5 &  &  \\
ZTF18aahhpyk & 142.55552 & 49.48830 & 2 & AGN &  &  & 1 &  &  \\
ZTF18aahhvnr & 192.25703 & 28.14522 & 4 & NT &  &  & 1 &  &  \\
ZTF18aahitda & 165.28990 & 51.36865 & 573 & AGN &  &  & 1 & Bronze & 0.252 \\
ZTF18aahjcqm & 209.15875 & 41.69028 & 13 & NT &  &  & 1,2 &  &  \\
ZTF18aahlyfs & 176.60789 & 11.12072 & 2 & NT &  &  & 1 &  &  \\
ZTF18aahpyvk & 158.09435 & 40.16202 & 4 & NT &  &  & 1 &  &  \\
ZTF18aahpyvz & 158.81903 & 39.89783 & 2 & NT &  &  & 1 &  &  \\
ZTF18aahruqt & 194.24561 & 47.15973 & 2 & NT &  &  & 1,2 &  &  \\
ZTF18aahrytm & 191.49862 & 40.77485 & 15 & NT &  &  & 1 &  &  \\
ZTF18aahsldb & 179.93421 & 25.62104 & 4 & NT &  &  & 1 &  &  \\
ZTF18aahugzi & 202.42121 & 36.90520 & 2 & AGN &  &  & 1 &  &  \\
ZTF18aahvlpf & 194.87449 & 31.33566 & 2 & NT &  &  & 1 &  &  \\
ZTF18aahvqfu & 199.61583 & 41.23478 & 3 & NT &  &  & 5,8 &  &  \\
ZTF18aaiaasw & 230.50746 & 43.53232 & 9 & NT &  &  & 1,2 &  &  \\
ZTF18aaierdy & 190.18980 & 49.99357 & 4 & NT &  &  & 1 &  &  \\
ZTF18aaigdlt & 170.53431 & 19.54339 & 3 & NT &  &  & 1,2 &  &  \\
ZTF18aaigdoo & 170.53412 & 19.65605 & 2 & NT &  &  & 5 &  &  \\
ZTF18aaiidgi & 191.61182 & 50.79206 & 7 & NT &  &  & 1,2 &  &  \\
ZTF18aaijdfe & 207.13159 & 26.94995 & 5 & NT &  &  & 1 &  &  \\
ZTF18aailgpx & 207.24917 & 57.28680 & 2 & NT &  &  & 1 &  &  \\
ZTF18aaitirs & 193.99900 & 27.95508 & 2 & NT &  &  & 1 &  &  \\
ZTF18aaitisu & 194.41423 & 27.60608 & 2 & NT &  &  & 5 &  &  \\
ZTF18aaititf & 194.44987 & 27.72655 & 2 & NT &  &  & 1 &  &  \\
ZTF18aaititz & 194.55010 & 27.12764 & 2 & NT &  &  & 1 &  &  \\
ZTF18aaitiuh & 194.89960 & 27.26185 & 4 & NT &  &  & 1 &  &  \\
ZTF18aaiuidq & 205.06954 & 30.13916 & 2 & NT &  &  & 1 &  &  \\
ZTF18aaiumcj & 207.87447 & 36.97036 & 4 & NT &  &  & 1 &  &  \\
ZTF18aaivfcz & 220.38522 & 46.67626 & 2 & NT &  &  & 1,2 &  &  \\
ZTF18aaizcpj & 227.22954 & 37.55827 & 2 & NT &  &  & 1 &  &  \\
ZTF18aajhtpx & 161.77724 & 29.21479 & 2 & NT &  &  & 1 &  &  \\
ZTF18aajiwps & 182.65186 & 37.09352 & 2 & NT &  &  & 1 &  &  \\
ZTF18aajjcxo & 179.82603 & 56.01073 & 3 & NT &  &  & 1 &  &  \\
ZTF18aakclqr & 169.97609 & 33.09011 & 2 & NT &  &  & 1 &  &  \\
ZTF18aakdcaa & 180.83726 & 32.47185 & 4 & NT &  &  & 1 &  &  \\
ZTF18aakearo & 203.34913 & 53.35837 & 3 & NT &  &  & 1,2 &  &  \\
ZTF18aakeljn & 207.45322 & 54.61891 & 9 & NT &  &  & 1 &  &  \\
ZTF18aakexgg & 179.82740 & 29.76499 & 2 & NT &  &  & 1 &  &  \\
ZTF18aakexzw & 173.41881 & 61.88803 & 6 & NT &  &  & 1 &  &  \\
ZTF18aakkvsz & 194.63185 & 28.46500 & 8 & AGN &  &  & 1 &  &  \\
ZTF18aakkxxq & 207.33247 & 26.66142 & 2 & NT &  &  & 1 &  &  \\
ZTF18aaklbpu & 194.63964 & 27.26884 & 2 & NT &  &  & 5 &  &  \\
ZTF18aaklbrg & 194.64070 & 27.67467 & 2 & NT &  &  & 1 &  &  \\
ZTF18aaklevm & 179.48817 & 26.64495 & 2 & NT &  &  & 1 &  &  \\
ZTF18aakmawv & 180.15680 & 24.03613 & 5 & NT &  &  & 1 &  &  \\
ZTF18aakocga & 182.02554 & 24.64104 & 4 & NT &  &  & 1 &  &  \\
ZTF18aakopca & 181.78593 & 25.50764 & 4 & NT &  &  & 1 &  &  \\
ZTF18aakqyen & 176.73658 & 55.48222 & 2 & NT &  &  & 1 &  &  \\
ZTF18aaktrha & 202.97896 & 47.88283 & 2 & NT &  &  & 1,2 &  &  \\
ZTF18aakycgz & 240.93534 & 52.40347 & 3 & NT &  &  & 1 &  &  \\
ZTF18aaloxok & 166.69943 & 24.92974 & 2 & NT &  &  & 1 &  &  \\
ZTF18aalpnoq & 172.95417 & 15.89748 & 2 & NT &  &  & 5 &  &  \\
ZTF18aalpymq & 196.35761 & 53.59176 & 9 & NT &  &  & 1,2 &  &  \\
ZTF18aalqeng & 184.26493 & 46.36008 & 2 & NT &  &  & 1 &  &  \\
ZTF18aalqgpx & 184.66142 & 44.53526 & 2 & NT &  &  & 1 &  &  \\
ZTF18aalvtaz & 169.37998 & 37.04431 & 2 & NT &  &  & 1,2 &  &  \\
ZTF18aamtgdp & 195.58759 & 47.63087 & 2 & NT &  &  & 1 &  &  \\
ZTF18aamtwrd & 215.59798 & 61.69753 & 9 & NT &  &  & 1 &  &  \\
ZTF18aamvcmk & 223.73102 & 45.52405 & 3 & NT &  &  & 1,2 &  &  \\
ZTF18aamvuds & 226.15446 & 48.73879 & 2 & NT &  &  & 1 &  &  \\
ZTF18aamzhyk & 233.73798 & 58.49971 & 2 & NT &  &  & 1 &  &  \\
ZTF18aanajij & 204.49158 & 65.73624 & 5 & NT &  &  & 1,2 &  &  \\
ZTF18aancdpi & 219.52642 & 30.50827 & 2 & NT &  &  & 1 &  &  \\
ZTF18aanyflw & 258.52783 & 57.99358 & 4 & NT &  &  & 5 &  &  \\
ZTF18aaobdql & 175.74104 & 54.95524 & 4 & NT &  &  & 1 &  &  \\
ZTF18aaodxwb & 192.30108 & 30.49497 & 2 & NT &  &  & 1 &  &  \\
ZTF18aaoszhl & 202.17625 & 55.36532 & 4 & NT & AT 2018kmp &  & 1 &  &  \\
ZTF18aaozhtf & 252.51060 & 26.42819 & 2 & NT & AT 2020hpd &  & 1 &  &  \\
ZTF18aaqccbr & 161.31841 & 39.38040 & 2 & NT &  &  & 1 &  &  \\
ZTF18aaqdfzb & 189.28210 & 10.70564 & 2 & NT &  &  & 1,5 &  &  \\
ZTF18aaqjlvs & 178.85462 & 56.74740 & 2 & NT &  &  & 1 &  &  \\
ZTF18aaqjuut & 197.79692 & 39.31070 & 2 & NT &  &  & 1 &  &  \\
ZTF18aaqjvng & 199.61664 & 32.53776 & 2 & NT &  &  & 1 &  &  \\
ZTF18aaqjxfc & 188.13463 & 40.01290 & 3 & NT &  &  & 1 &  &  \\
ZTF18aaqkryj & 188.58514 & 65.50501 & 3 & NT &  &  & 1 &  &  \\
ZTF18aaqldph & 199.80184 & 31.64632 & 2 & NT &  &  & 1 &  &  \\
ZTF18aaqmphu & 219.43947 & 51.40579 & 69 & NT &  &  & 1,2 & Bronze & 0.148 \\
ZTF18aaqqgit & 161.20284 & 51.95051 & 242 & NT & AT 2018ahh &  & 1 & Bronze & 0.064 \\
ZTF18aaqrssj & 218.44611 & 54.66503 & 2 & NT &  &  & 1 &  &  \\
ZTF18aaqsbte & 238.66114 & 55.91026 & 2 & NT &  &  & 1 &  &  \\
ZTF18aarbklo & 158.63712 & 19.22056 & 8 & NT &  &  & 1 &  &  \\
ZTF18aarcjhu & 181.14671 & 30.10879 & 2 & NT &  &  & 1 &  &  \\
ZTF18aardnpw & 165.15395 & 44.09749 & 2 & NT &  &  & 1 &  &  \\
ZTF18aarfvib & 233.69244 & 31.57719 & 2 & NT &  &  & 1 &  &  \\
ZTF18aaricta & 256.48927 & 63.01544 & 2 & NT &  &  & 1,2 &  &  \\
ZTF18aariyxn & 207.02868 & 26.69441 & 2 & NT &  &  & 1 &  &  \\
ZTF18aarlbsg & 176.73969 & 31.21330 & 2 & NT &  &  & 1,2 &  &  \\
ZTF18aarlbvb & 176.72379 & 31.02253 & 2 & NT &  &  & 1 &  &  \\
ZTF18aarlbvk & 175.92728 & 30.64308 & 2 & NT &  &  & 1 &  &  \\
ZTF18aarlieh & 201.56987 & 58.68326 & 2 & AGN &  &  & 1 &  &  \\
ZTF18aarlqcb & 179.38820 & 32.61041 & 2 & NT &  &  & 1 &  &  \\
ZTF18aarzxlx & 201.31726 & 52.45665 & 5 & NT &  &  & 1 &  &  \\
ZTF18aastwrz & 180.51064 & 15.01072 & 6 & AGN & AT 2018bvy &  & 1 &  &  \\
ZTF18aasvckm & 197.94258 & 14.57041 & 2 & NT &  &  & 5 &  &  \\
ZTF18aasxihq & 197.96059 & 19.34811 & 84 & AGN &  &  & 1 & Bronze & 0.398 \\
ZTF18aasxikm & 197.96059 & 19.34811 & 2 & AGN &  &  & 1 &  &  \\
ZTF18aasyrnk & 219.41440 & 29.28828 & 7 & NT &  &  & 5 &  &  \\
ZTF18aaszvlm & 221.06435 & 55.37334 & 2 & NT &  &  & 1 &  &  \\
ZTF18aathzlp & 218.72835 & 8.05178 & 2 & NT &  &  & 1,2 &  &  \\
ZTF18aauvezc & 166.83199 & 46.38329 & 2 & NT &  &  & 5 &  &  \\
ZTF18aauxynp & 221.28056 & 52.15134 & 2 & NT &  &  & 1,2 &  &  \\
ZTF18aavdqoy & 278.84087 & 53.45010 & 103 & VS &  &  & 5 & Bronze &  \\
ZTF18aaviokz & 207.86833 & 40.44796 & 12 & NT & AT 2021duz &  & 1 &  &  \\
ZTF18aavrofn & 165.17610 & 40.34667 & 2 & NT &  &  & 1 &  &  \\
ZTF18aawamsd & 299.38607 & 36.07132 & 6 & VS &  &  & 5,8 &  &  \\
ZTF18aawjgxa & 247.47959 & 24.02946 & 2 & NT &  &  & 1 &  &  \\
ZTF18aawmnql & 135.30920 & 51.20681 & 17 & NT &  &  & 1 &  &  \\
ZTF18aawnjqz & 167.62140 & 30.69830 & 2 & NT &  &  & 5 &  &  \\
ZTF18aawnqoc & 171.99696 & 35.77907 & 2 & NT &  &  & 1 &  &  \\
ZTF18aawohpc & 177.52612 & 26.58845 & 2 & NT &  &  & 1 &  &  \\
ZTF18aawolyo & 154.21407 & 22.49742 & 2 & NT &  &  & 1 &  &  \\
ZTF18aawpght & 169.73594 & 23.66164 & 2 & AGN &  &  & 1 &  &  \\
ZTF18aawqupi & 229.53917 & 35.85220 & 4 & NT &  &  & 1 &  &  \\
ZTF18aawxmbu & 172.67019 & 56.48612 & 2 & NT &  &  & 1,2 &  &  \\
ZTF18aaxeoqk & 216.65238 & 15.40177 & 8 & NT &  &  & 1 &  &  \\
ZTF18aaxlucv & 298.53403 & 21.56921 & 19 & VS &  &  & 5 &  &  \\
ZTF18aaxqtyr & 156.72910 & 45.69590 & 2 & NT &  &  & 1 &  &  \\
ZTF18aaxzlva & 172.15612 & 50.82742 & 7 & NT &  &  & 1 &  &  \\
ZTF18aaybewz & 214.32642 & 46.69823 & 2 & NT &  &  & 1 &  &  \\
ZTF18aaybkrg & 262.05487 & 51.26042 & 17 & SN &  &  & 5 &  &  \\
ZTF18aayijie & 195.59515 & 31.96619 & 3 & NT &  &  & 5 &  &  \\
ZTF18aayimoj & 196.32216 & 32.17940 & 2 & NT &  &  & 1 &  &  \\
ZTF18aayinmz & 183.41029 & 40.33776 & 2 & AGN &  &  & 1 &  &  \\
ZTF18aazjsus & 222.49990 & 49.86213 & 2 & NT &  &  & 1 &  &  \\
ZTF18aazogjs & 200.51138 & 25.48533 & 3 & NT &  &  & 5 &  &  \\
ZTF18abaeuzc & 158.76575 & 52.15126 & 40 & NT &  &  & 1 & Bronze & 0.143 \\
ZTF18abakxep & 186.84405 & 30.09254 & 17 & NT & AT 2019dec &  & 1 &  &  \\
ZTF18abalpqy & 183.13028 & 15.27034 & 6 & AGN &  &  & 1 &  &  \\
ZTF18abawbfn & 227.65716 & 61.74347 & 3 & NT &  &  & 1 &  &  \\
ZTF18abbiyrc & 249.29601 & 25.45319 & 3 & NT &  &  & 1 &  &  \\
ZTF18abbjazr & 232.32219 & 30.87345 & 2 & NT &  &  & 1 &  &  \\
ZTF18abbmmqp & 284.48631 & 29.20436 & 4 & NT &  &  & 5 &  &  \\
ZTF18abdmqna & 218.83266 & 30.09610 & 2 & NT &  &  & 1 &  &  \\
ZTF18abeaizq & 241.56769 & 21.06364 & 5 & NT &  &  & 5 &  &  \\
ZTF18abeakzq & 239.22698 & 25.54267 & 18 & AGN &  &  & 1 &  &  \\
ZTF18abguhwj & 173.80969 & 45.03930 & 4 & NT &  &  & 1 &  &  \\
ZTF18abhaejc & 355.91228 & 54.50881 & 2 & VS &  &  & 5 &  &  \\
ZTF18abjpdmt & 16.65585 & 58.54964 & 66 & VS &  &  & 5 & Bronze &  \\
ZTF18abjyiua & 235.22248 & 9.74992 & 2 & NT &  &  & 1 &  &  \\
ZTF18abkefdc & 336.05717 & 32.34489 & 2 & VS &  &  & 5 &  &  \\
ZTF18ablpfdj & 274.47118 & 5.82349 & 13 & VS &  &  & 5,8 &  &  \\
ZTF18abmhkff & 297.39815 & 13.01845 & 2 & VS &  &  & 5 &  &  \\
ZTF18abmrhom & 325.04727 & 21.55846 & 13 & NT & SN 2018ffi & SN Ia & 5 &  &  \\
ZTF18abmrlqs & 329.47782 & 55.16463 & 5 & VS &  &  & 8 &  &  \\
ZTF18abmwycg & 233.82758 & 26.64937 & 2 & NT &  &  & 1 &  &  \\
ZTF18abnvgif & 24.63887 & 54.75873 & 2 & NT &  &  & 5 &  &  \\
ZTF18aboswes & 24.48174 & 56.08004 & 4 & SN &  &  & 5 &  &  \\
ZTF18abottlo & 17.40277 & -15.98532 & 4 & NT &  &  & 6 &  &  \\
ZTF18abpeoqr & 20.35424 & -16.01233 & 3 & VS &  &  & 6 &  &  \\
ZTF18abqjfnx & 240.11169 & 53.25051 & 2 & NT &  &  & 1 &  &  \\
ZTF18abrgace & 287.00244 & -8.07820 & 4 & VS &  &  & 5 &  &  \\
ZTF18abryusp & 34.40855 & 29.74434 & 13 & VS &  &  & 5 &  &  \\
ZTF18absgoav & 350.16196 & 42.86538 & 2 & NT &  &  & 5 &  &  \\
ZTF18abtafdw & 321.34911 & 55.98211 & 4 & SN &  &  & 5 &  &  \\
ZTF18abteulk & 330.57046 & 52.49530 & 2 & VS &  &  & 5 &  &  \\
ZTF18abtfwhe & 15.17917 & -0.48611 & 4 & NT &  &  & 5 &  &  \\
ZTF18abtgunq & 38.44556 & -1.02454 & 279 & AGN & AT 2018cqh &  & 1 & Silver & 0.049 \\
ZTF18abtmtit & 36.64124 & -1.10785 & 231 & NT &  &  & 1 & Bronze & 0.096 \\
ZTF18abtsxba & 34.77290 & -17.42028 & 10 & AGN &  &  & 6 &  &  \\
ZTF18abttdtx & 97.60642 & 63.67812 & 15 & AGN &  &  & 5 &  &  \\
ZTF18abugoat & 332.14588 & 55.09960 & 13 & VS &  &  & 5 &  &  \\
ZTF18abupbwb & 27.78328 & -7.53520 & 10 & NT &  &  & 6 &  &  \\
ZTF18abupcpv & 28.82121 & -6.18079 & 3 & NT &  &  & 6 &  &  \\
ZTF18abvttfg & 137.03671 & 50.15561 & 8 & NT &  &  & 1 &  &  \\
ZTF18abwmuua & 249.64342 & 31.87296 & 4 & NT & SN 2018gvb & SN Ia & 1 &  &  \\
ZTF18abwnzru & 305.44510 & 33.88024 & 3 & VS &  &  & 5 &  &  \\
ZTF18abxmzjw & 26.52993 & 23.44960 & 2 & SN &  &  & 5 &  &  \\
ZTF18abxpfzq & 21.75555 & 0.94317 & 2 & NT &  &  & 6 &  &  \\
ZTF18abxplzo & 317.17288 & 54.14834 & 5 & VS &  &  & 5 &  &  \\
ZTF18abxrrtb & 40.18780 & -20.61313 & 3 & NT &  &  & 6 &  &  \\
ZTF18abxrsex & 32.08180 & -7.18915 & 9 & NT &  &  & 6 &  &  \\
ZTF18abxrxwj & 27.27226 & -12.21063 & 13 & NT &  &  & 6 &  &  \\
ZTF18abxrzrl & 26.62387 & -3.91350 & 2 & NT &  &  & 6 &  &  \\
ZTF18abxtzvo & 52.46168 & -13.38191 & 6 & NT &  &  & 5 &  &  \\
ZTF18abxudzf & 26.20205 & -6.18832 & 2 & NT &  &  & 5 &  &  \\
ZTF18acalhmo & 26.60055 & -15.86588 & 2 & SN &  &  & 6 &  &  \\
ZTF18acbxqdr & 354.67484 & -10.71099 & 2 & NT &  &  & 1 &  &  \\
ZTF18accvowz & 169.60798 & 36.87661 & 2 & VS &  &  & 5 &  &  \\
ZTF18accvpqx & 170.04222 & 38.10219 & 2 & NT &  &  & 1 &  &  \\
ZTF18acdmrhz & 350.00448 & 55.48388 & 7 & VS &  &  & 5 &  &  \\
ZTF18aceajmw & 60.78337 & -5.76494 & 2 & NT &  &  & 1 &  &  \\
ZTF18acepkas & 136.84486 & 44.11084 & 2 & NT &  &  & 1 &  &  \\
ZTF18acerrai & 174.62396 & 32.17043 & 6 & NT &  &  & 1 &  &  \\
ZTF18acfvwnv & 40.04359 & -12.50032 & 2 & SN &  &  & 6 &  &  \\
ZTF18acgunuw & 178.93276 & 80.21923 & 3 & NT &  &  & 5,8 &  &  \\
ZTF18acgwntr & 147.59309 & 25.79599 & 2 & NT &  &  & 5,8 &  &  \\
ZTF18achzddr & 123.32063 & 22.64830 & 6 & NT & AT 2019azh & TDE & 1,2 &  &  \\
ZTF18aciblji & 174.01353 & 17.35970 & 2 & NT &  &  & 1 &  &  \\
ZTF18acidntq & 126.61549 & 8.74349 & 21 & AGN & AT 2019ofz &  & 1 & Bronze & 0.082 \\
ZTF18acjvwsu & 10.41444 & 1.06842 & 5 & NT &  &  & 1,2,6 &  &  \\
ZTF18acmtyar & 94.94340 & 19.15672 & 4 & VS &  &  & 5 &  &  \\
ZTF18acpdhit & 140.95394 & 24.81479 & 5 & NT &  &  & 1 &  &  \\
ZTF18acpdvos & 151.71195 & 1.69278 & 92 & NT & AT 2018hyz & TDE & 1,2 &  &  \\
ZTF18acpeeih & 196.08742 & 1.15098 & 26 & NT & AT 2018ail & Galaxy & 1 &  &  \\
ZTF18acpljng & 140.28578 & 44.91415 & 77 & AGN &  &  & 1 & Bronze & 0.156 \\
ZTF18acpmhuj & 170.08733 & 32.24833 & 2 & NT &  &  & 1 &  &  \\
ZTF18acpntil & 153.16353 & 46.52167 & 3 & NT &  &  & 1 &  &  \\
ZTF18acpoxaq & 116.43683 & 46.25113 & 9 & NT &  &  & 1,2 &  &  \\
ZTF18acqeuaj & 131.11845 & 19.09025 & 2 & NT &  &  & 1 &  &  \\
ZTF18acqptsi & 213.99632 & 38.89357 & 2 & NT &  &  & 1 &  &  \\
ZTF18acqyart & 126.61549 & 8.74349 & 38 & AGN & AT 2019ofz &  & 1 & Bronze & 0.082 \\
ZTF18acrmldj & 152.56724 & -0.07478 & 2 & NT &  &  & 1,2 &  &  \\
ZTF18acsoyiv & 153.45777 & 39.64187 & 3 & NT & AT 2018jkl &  & 1 &  &  \\
ZTF18acsremz & 148.62736 & 53.16918 & 49 & NT & SN 2021hmc & SN Ia & 1 &  &  \\
ZTF18acsripv & 151.42136 & 37.62661 & 2 & NT &  &  & 1 &  &  \\
ZTF18acusldo & 156.19250 & 10.91342 & 2 & NT &  &  & 1 &  &  \\
ZTF18acvgzlg & 198.24895 & 18.17804 & 44 & VS &  &  & 5 & Bronze &  \\
ZTF18acvimfo & 223.28041 & 3.53817 & 3 & AGN &  &  & 1 &  &  \\
ZTF18acybdqr & 170.35937 & 28.23510 & 2 & NT &  &  & 1 &  &  \\
ZTF18aczenvx & 163.73024 & 27.80237 & 12 & NT &  &  & 1 &  &  \\
ZTF18adbditf & 16.65585 & 58.54964 & 43 & VS &  &  & 5 & Bronze &  \\
ZTF18adblgvo & 305.14883 & 19.80043 & 2 & SN &  &  & 5 &  &  \\
ZTF18adcassp & 241.98193 & 9.46367 & 2 & NT &  &  & 1 &  &  \\
ZTF19aaabslc & 14.14184 & -3.34520 & 2 & NT &  &  & 5 &  &  \\
ZTF19aaadfcp & 120.01926 & 58.76128 & 6 & SN & AT 2019gq &  & 5 &  &  \\
ZTF19aaakdpz & 78.52702 & 2.51732 & 4 & NT &  &  & 5 &  &  \\
ZTF19aabpdck & 110.59545 & -10.43971 & 2 & VS &  &  & 5 &  &  \\
ZTF19aabybwz & 127.48938 & 44.94016 & 4 & NT &  &  & 1 &  &  \\
ZTF19aacsofi & 120.01938 & 40.07791 & 7 & NT & SN 2019oc & SN Ia & 1 &  &  \\
ZTF19aadufvo & 268.54761 & 13.90573 & 4 & NT &  &  & 5 &  &  \\
ZTF19aadymum & 200.74947 & 27.11643 & 3 & NT &  &  & 1 &  &  \\
ZTF19aafncky & 240.76515 & 21.69747 & 2 & NT &  &  & 1,2 &  &  \\
ZTF19aafzmzl & 132.92260 & 48.57251 & 2 & NT &  &  & 1 &  &  \\
ZTF19aagrxjd & 140.77516 & 24.02489 & 2 & NT &  &  & 1,2 &  &  \\
ZTF19aakmdrn & 37.43868 & 53.88640 & 2 & VS &  &  & 5,8 &  &  \\
ZTF19aalfugu & 248.50624 & 77.60058 & 3 & NT & AT 2019mds &  & 5 &  &  \\
ZTF19aamhhgu & 222.97147 & 55.40781 & 22 & NT & SN 2019bxh & SN Ia & 1 &  &  \\
ZTF19aamohnt & 192.01674 & 18.97596 & 2 & NT & AT 2019bvk &  & 1 &  &  \\
ZTF19aamrjve & 250.26490 & 32.26472 & 2 & NT &  &  & 1,2 &  &  \\
ZTF19aanleed & 231.77844 & 34.06871 & 2 & NT & AT 2020vgn &  & 1 &  &  \\
ZTF19aaoxijx & 161.54037 & 24.26686 & 9 & NT & AT 2019cxz &  & 1 &  &  \\
ZTF19aapatmf & 258.89200 & 36.17269 & 2 & SN &  &  & 5 &  &  \\
ZTF19aarepdu & 88.35785 & 11.38558 & 99 & NT & AT 2019aale & AGN & 5 & Bronze &  \\
ZTF19aarsqmc & 120.73166 & 67.93110 & 5 & UNCLEAR &  &  & 5 &  &  \\
ZTF19aavlnpb & 211.64991 & 8.75663 & 2 & NT &  &  & 1,2 &  &  \\
ZTF19aavocqz & 221.02966 & 18.01263 & 5 & NT &  &  & 1 &  &  \\
ZTF19aavowrh & 246.27715 & 46.48351 & 5 & NT &  &  & 1 &  &  \\
ZTF19aayozuj & 230.99805 & 25.97714 & 4 & NT &  &  & 1 &  &  \\
ZTF19aayrzba & 320.87642 & -7.74598 & 12 & NT &  &  & 1 &  &  \\
ZTF19aazylqj & 326.96732 & 45.48559 & 4 & VS &  &  & 5 &  &  \\
ZTF19abaevrx & 156.56979 & 53.41965 & 7 & NT &  &  & 1 &  &  \\
ZTF19abakysz & 234.26447 & 10.50281 & 2 & NT &  &  & 1,2 &  &  \\
ZTF19abboojm & 316.42663 & -5.62293 & 2 & NT &  &  & 1 &  &  \\
ZTF19abcupln & 242.67822 & 15.15571 & 12 & NT & SN 2019kcj & SN Ia & 1 &  &  \\
ZTF19abdkfbr & 24.81343 & 2.01678 & 95 & NT &  &  & 6 & Bronze &  \\
ZTF19abdsrof & 159.24873 & 25.92185 & 4 & NT &  &  & 1 &  &  \\
ZTF19abeyuen & 319.63187 & 10.11603 & 13 & NT &  &  & 1 &  &  \\
ZTF19abgwjfa & 298.35551 & 70.35697 & 5 & VS &  &  & 5 &  &  \\
ZTF19abixawb & 2.56205 & 0.13911 & 20 & AGN &  &  & 1 & Bronze & 0.102 \\
ZTF19abktbzk & 246.29572 & 46.44702 & 3 & NT &  &  & 1 &  &  \\
ZTF19abmonpc & 352.73518 & 13.83201 & 7 & NT &  &  & 1 &  &  \\
ZTF19abnktav & 249.48949 & 11.60236 & 6 & NT &  &  & 1 &  &  \\
ZTF19abofgnr & 305.79420 & 15.39329 & 3 & VS &  &  & 5 &  &  \\
ZTF19abpuikr & 237.65588 & 23.04692 & 5 & NT & SN 2019nvh & SN Ia & 1 &  &  \\
ZTF19absrfvd & 352.59499 & 32.49691 & 2 & VS &  &  & 5 &  &  \\
ZTF19abssxkv & 341.57856 & 37.14076 & 2 & VS &  &  & 5,8 &  &  \\
ZTF19abtladp & 43.82673 & 20.12041 & 2 & VS &  &  & 5 &  &  \\
ZTF19abtrvsq & 28.43509 & -23.61405 & 9 & NT &  &  & 6 &  &  \\
ZTF19abuoqzz & 339.86966 & 16.62459 & 45 & NT &  &  & 5 & Bronze &  \\
ZTF19abxbybm & 285.89145 & -16.15508 & 2 & VS &  &  & 5 &  &  \\
ZTF19abymgox & 58.46971 & -26.41953 & 26 & VS &  &  & 5,8 & Bronze &  \\
ZTF19abymuda & 97.10032 & -17.28794 & 36 & SN & AT 2019nks &  & 5 & Bronze &  \\
ZTF19abzlmxk & 84.34894 & 51.63366 & 30 & NT &  &  & 5,8 & Bronze &  \\
ZTF19acaheuq & 29.05425 & 1.03518 & 2 & NT &  &  & 1 &  &  \\
ZTF19acajpme & 222.22615 & 31.65806 & 2 & NT & AT 2019rvr &  & 1,2 &  &  \\
ZTF19acbkqdn & 320.21980 & 11.12030 & 5 & NT &  &  & 1 &  &  \\
ZTF19acbmotn & 117.01818 & 22.80831 & 20 & NT & AT 2019rru &  & 1 & Gold & 0.075 \\
ZTF19ackilif & 121.39812 & 24.48656 & 3 & NT &  &  & 1 &  &  \\
ZTF19acudnoy & 342.21106 & -0.49405 & 2 & NT &  &  & 1 &  &  \\
ZTF19acuhpoi & 0.86881 & 0.45835 & 2 & NT &  &  & 1 &  &  \\
ZTF19acumsmk & 34.09070 & -2.94162 & 2 & NT &  &  & 6 &  &  \\
ZTF19acvtuva & 155.16052 & -3.72954 & 117 & NT & AT 2020zet &  & 5 & Silver & 0.173 \\
ZTF19acxgwid & 346.50389 & 0.31121 & 9 & NT & SN 2019vva & SN Ia & 1 &  &  \\
ZTF19acxlcdz & 124.39072 & 56.04858 & 39 & NT & SN 2019vtx & SN Ia & 1 &  &  \\
ZTF19acxyvwp & 170.21654 & -1.22918 & 2 & NT &  &  & 1 &  &  \\
ZTF19aczjwxq & 148.11244 & 36.71754 & 3 & NT & AT 2019xeg &  & 1 &  &  \\
ZTF19adakvxy & 325.53655 & -7.33822 & 2 & NT & AT 2019xdn &  & 1,2 &  &  \\
ZTF19adbwqgp & 121.65541 & 25.43951 & 2 & NT &  &  & 5 &  &  \\
ZTF19adcdxlv & 169.88449 & 18.69028 & 46 & NT &  &  & 8 & Bronze &  \\
ZTF20aaapfih & 115.67725 & 40.35026 & 3 & NT &  &  & 1 &  &  \\
ZTF20aaawbkz & 147.52718 & 5.72373 & 10 & NT & SN 2020K & SN Ia & 1 &  &  \\
ZTF20aabpmlx & 157.73044 & -19.45966 & 4 & NT & AT 2020as &  & 5 &  &  \\
ZTF20aadxwuh & 182.45354 & 22.74239 & 32 & AGN &  &  & 1 & Bronze & 0.026 \\
ZTF20aaenbrp & 158.53777 & 4.35856 & 3 & NT &  &  & 1 &  &  \\
ZTF20aaenpjf & 169.59660 & 3.43240 & 2 & NT &  &  & 1 &  &  \\
ZTF20aaerplp & 227.14104 & 25.48306 & 2 & NT &  &  & 1 &  &  \\
ZTF20aaerpyl & 228.53823 & 32.67463 & 3 & AGN &  &  & 1 &  &  \\
ZTF20aafcjln & 182.90434 & 0.32250 & 42 & NT & SN 2020vf & SN Ia & 1 &  &  \\
ZTF20aafjjfv & 134.13665 & 35.10125 & 2 & NT &  &  & 1 &  &  \\
ZTF20aagffdu & 154.10567 & 42.09316 & 17 & NT & AT 2020abg &  & 1 &  &  \\
ZTF20aagiirb & 237.51859 & 39.25033 & 8 & NT &  &  & 1 &  &  \\
ZTF20aagoidp & 355.28357 & 49.83056 & 2 & VS &  &  & 5 &  &  \\
ZTF20aagxiaq & 170.87678 & 29.22858 & 2 & NT &  &  & 1 &  &  \\
ZTF20aahgpaj & 164.55025 & 11.88881 & 2 & NT &  &  & 1 &  &  \\
ZTF20aahjzhk & 131.86646 & 3.66901 & 2 & AGN &  &  & 1 &  &  \\
ZTF20aahmzxd & 112.52822 & -5.41436 & 59 & AGN &  &  & 5,8 & Bronze &  \\
ZTF20aajbdvu & 132.65559 & 39.45760 & 5 & NT &  &  & 1 &  &  \\
ZTF20aajccqq & 229.53817 & 45.17101 & 16 & NT & AT 2020bwk &  & 7 &  &  \\
ZTF20aajoyjf & 123.50894 & 21.36742 & 18 & NT & AT 2020bub &  & 1 &  &  \\
ZTF20aaozcrd & 192.59831 & -4.25625 & 11 & NT &  &  & 5,8 &  &  \\
ZTF20aaqsrdk & 177.63922 & 17.85810 & 2 & NT &  &  & 1 &  &  \\
ZTF20aaqtmxr & 232.53647 & 18.81420 & 5 & NT &  &  & 5,8 &  &  \\
ZTF20aaraxda & 180.71940 & 27.17761 & 2 & NT &  &  & 5 &  &  \\
ZTF20aatpsar & 190.93488 & 53.31267 & 2 & NT &  &  & 1,2 &  &  \\
ZTF20aauolxq & 229.41904 & 13.85171 & 11 & NT & AT 2020fwm &  & 1,2 &  &  \\
ZTF20aavevtg & 130.23473 & 10.80814 & 2 & NT &  &  & 1 &  &  \\
ZTF20aavtydb & 119.34340 & 52.60993 & 2 & NT &  &  & 1 &  &  \\
ZTF20aavynba & 208.28754 & 47.29955 & 18 & NT & SN 2020ism & SN Ia & 1 &  &  \\
ZTF20aawjeyc & 146.51510 & 31.73927 & 5 & NT &  &  & 5 &  &  \\
ZTF20aawjwkh & 136.80276 & 15.60691 & 2 & NT &  &  & 1 &  &  \\
ZTF20aayngca & 138.76935 & 10.25877 & 6 & NT & SN 2020jfn & SN Ia & 1 &  &  \\
ZTF20aazfhyf & 331.24184 & 1.00129 & 3 & NT & AT 2020kxs &  & 1 &  &  \\
ZTF20aazgtmp & 180.74383 & 20.08520 & 20 & NT & SN 2020jny & SN Ia & 1 &  &  \\
ZTF20aazmsko & 147.48496 & -0.23136 & 2 & NT &  &  & 1 &  &  \\
ZTF20abblwrx & 234.53291 & 46.68459 & 2 & NT &  &  & 1 &  &  \\
ZTF20abbvaqi & 261.04990 & 31.77627 & 5 & NT &  &  & 1,2 &  &  \\
ZTF20abfrbyv & 155.54340 & 80.20390 & 2 & VS &  &  & 5 &  &  \\
ZTF20abhuzpg & 233.94868 & 24.57072 & 2 & NT &  &  & 1 &  &  \\
ZTF20abidglv & 42.66854 & 41.67134 & 3 & NT &  &  & 5 &  &  \\
ZTF20abimsxj & 252.52541 & 32.30181 & 14 & NT & AT 2020vao &  & 1 &  &  \\
ZTF20abismcv & 163.71303 & 33.15199 & 2 & NT &  &  & 1 &  &  \\
ZTF20abkxrun & 183.02606 & 40.82678 & 4 & NT &  &  & 1 &  &  \\
ZTF20ablgxjx & 235.15980 & 47.55920 & 3 & NT &  &  & 1 &  &  \\
ZTF20abmhfza & 260.79061 & 61.37880 & 2 & NT &  &  & 1 &  &  \\
ZTF20aboquqi & 8.90010 & 31.67528 & 8 & VS &  &  & 5 &  &  \\
ZTF20abqosnh & 23.97295 & 39.95597 & 70 & SN & SN 2020rba & SN Ia & 5 &  &  \\
ZTF20abrclgl & 11.26834 & 17.69454 & 33 & SN &  &  & 5 & Silver &  \\
ZTF20abtjhgf & 31.43169 & 19.16404 & 40 & SN &  &  & 5 & Bronze &  \\
ZTF20abujvgp & 49.25103 & 40.72814 & 2 & NT &  &  & 5 &  &  \\
ZTF20abxphdt & 39.25395 & 26.62206 & 45 & SN & AT 2020skf &  & 5 & Gold & 0.068 \\
ZTF20abzvzkn & 341.00105 & 39.13234 & 9 & SN &  &  & 5 &  &  \\
ZTF20acaecko & 354.30551 & -10.09510 & 2 & NT &  &  & 1,2 &  &  \\
ZTF20acbeztg & 22.15925 & -17.21419 & 28 & VS &  &  & 6,9 & Bronze &  \\
ZTF20achuvja & 75.33495 & -22.19069 & 53 & NT &  &  & 6 & Bronze &  \\
ZTF20acimzuq & 115.75596 & 41.71005 & 9 & NT &  &  & 1 &  &  \\
ZTF20acitpfz & 136.35777 & 61.80255 & 30 & NT & AT 2020wey & TDE & 1 &  &  \\
ZTF20ackhzba & 142.20816 & 46.65336 & 27 & NT & SN 2020xsr & SN Ia & 1 &  &  \\
ZTF20ackryyi & 150.52813 & 10.61114 & 2 & AGN &  &  & 5 &  &  \\
ZTF20aclflri & 93.25657 & 70.34562 & 9 & SN & AT 2020xna &  & 5 &  &  \\
ZTF20aclgfji & 138.30025 & 5.13455 & 18 & NT & AT 2020ygl &  & 1 &  &  \\
ZTF20aclzyyl & 9.82111 & -17.20901 & 2 & NT &  &  & 5,8 &  &  \\
ZTF20acselme & 134.13635 & 39.14106 & 6 & NT &  &  & 1 &  &  \\
ZTF20adafhdh & 158.59053 & 16.46315 & 6 & NT &  &  & 1 &  &  \\
ZTF21aaaqmuf & 1.42153 & 4.13448 & 3 & SN & AT 2021ee &  & 6 &  &  \\
ZTF21aaaqxuf & 4.53661 & 2.51476 & 2 & NT &  &  & 5,8 &  &  \\
ZTF21aabasub & 30.22086 & -4.77991 & 14 & NT &  &  & 5 &  &  \\
ZTF21aacdjdi & 78.70575 & 48.03341 & 11 & SN & AT 2021rw &  & 5 &  &  \\
ZTF21aacilcv & 156.72314 & 0.55806 & 4 & NT &  &  & 1 &  &  \\
ZTF21aadmkrm & 79.57394 & 58.44900 & 17 & SN & AT 2021abe &  & 5 &  &  \\
ZTF21aagjlpo & 166.83199 & 46.38329 & 3 & NT &  &  & 5 &  &  \\
ZTF21aahfizx & 225.30988 & 13.02201 & 10 & NT & AT 2021bwg &  & 1 &  &  \\
ZTF21aahfjbs & 229.39751 & 18.11808 & 30 & NT & SN 2021cky & SN Ia & 1 &  &  \\
ZTF21aahuvvx & 82.80987 & -18.43544 & 2 & SN &  &  & 6 &  &  \\
ZTF21aaiahsu & 165.25533 & 8.08728 & 29 & NT &  &  & 1 & Gold & 0.088 \\
ZTF21aakjhgt & 250.27260 & 25.48488 & 2 & NT &  &  & 1 &  &  \\
ZTF21aambjvl & 114.78316 & 41.75661 & 3 & NT &  &  & 5 &  &  \\
ZTF21aandeyw & 224.67104 & 9.18540 & 2 & NT &  &  & 1 &  &  \\
ZTF21aaqhung & 176.49295 & 19.64975 & 2 & NT &  &  & 1,2 &  &  \\
ZTF21aaridnv & 203.02324 & 1.59735 & 2 & NT &  &  & 1 &  &  \\
ZTF21aatrajq & 217.69093 & 57.08014 & 2 & NT &  &  & 1 &  &  \\
ZTF21aauufbz & 160.22916 & 24.74712 & 6 & NT &  &  & 1 &  &  \\
ZTF21aaxxjen & 166.41431 & 19.46035 & 22 & NT & SN 2021kun & SN Ia & 1 &  &  \\
ZTF21aaxyfzb & 215.61877 & 31.75940 & 8 & NT & AT 2021lml &  & 1 &  &  \\
ZTF21aaycwpu & 123.96885 & 29.80591 & 8 & NT & AT 2021kxv &  & 1 &  &  \\
ZTF21aaydzrb & 200.83530 & -1.72122 & 12 & NT &  &  & 1 &  &  \\
ZTF21aazmjaf & 316.71008 & 10.73577 & 7 & NT & AT 2021lkq &  & 1 &  &  \\
ZTF21abbsncs & 233.68031 & 56.99315 & 2 & NT & AT 2021mkd &  & 1 &  &  \\
ZTF21abcpqpv & 216.72060 & 23.16110 & 4 & NT & AT 2021nzg &  & 1 &  &  \\
ZTF21abebkok & 330.12547 & 21.05209 & 5 & SN & AT 2021osm &  & 7 &  &  \\
ZTF21abhqqwq & 196.26842 & 60.77196 & 7 & VS & SN 2021qus & SN Ia & 7 &  &  \\
\hline

\end{longtable}

\tablecomments{$^a$``Sherlock''\footnote{\url{https://lasair.roe.ac.uk/sherlock}} is a contextual classifier which provides an initial classification of every non-moving object by performing a spatial cross-match against a large number of catalogs such as nearby galaxies, variable stars and active galactic nuceli (see Section 4.2 of \citealt{Smith2020} for more details).  AGN = Active Galactic Nucleus; NT = Nuclear Transient (i.e. a transient consistent with the nucleus of a galaxy); SN = Supernova; VS = Variable Star.}

\end{document}